\documentclass[11pt]{article}

\usepackage{amssymb}

\usepackage{graphicx}

\usepackage[english,francais]{babel}

\newfont{\ensmathquatorze}{msbm10 scaled 1400}
\newfont{\ensmathonze}{msbm10 scaled 1100}
\newfont{\ensmathdix}{msbm10}
\newfont{\ensmathneuf}{msbm10 scaled 833}
\newfont{\ensmathhuit}{msbm10 scaled 694}
\newfam\ensmathfam                        
\textfont\ensmathfam=\ensmathonze        
\scriptfont\ensmathfam=\ensmathdix       
\scriptscriptfont\ensmathfam=\ensmathhuit
\def\ensmf{\fam\ensmathfam\ensmathonze}         

\def\be{\begin{equation}}
\def\ee{\end{equation}}

\def\bea{\begin{eqnarray}}
\def\eea{\end{eqnarray}}
\def\beann{\begin{eqnarray*}}
\def\eeann{\end{eqnarray*}}

\def\typ{\mbox{\tiny typ}}

\renewcommand{\geq}{\geqslant}

\newcommand{\ket}[1]{|\kern.3ex#1\kern.3ex\rangle}
\newcommand{\bra}[1]{\langle\kern.3ex #1 \kern.3ex|}
\newcommand{\APPROX}[1]{                
   {{\raisebox{-.3cm}{$\textstyle\simeq$}} \atop {\scriptstyle{#1}}}}

\newcommand{\mean}[1]{\left\langle #1 \right\rangle} 
\newcommand{\smean}[1]{\langle #1 \rangle} 
\newcommand{\cum}[1]{\langle\langle #1 \rangle\rangle} 

\newcommand{\EXP}[1]{{\mbox{\large e}}^{#1}}         

\renewcommand{\min}[2]{\mathop{\mathrm{min}}\nolimits\left( #1 , #2\right)}

\newcommand{\proba}{\mathop{\mathrm{Prob}}\nolimits}  
\newcommand{\heav}{\mathop{\theta}\nolimits}  

\def\NN{{\ensmf N}}                 

\def\R{{\sf I\kern-.15em R}}
\def\C{\kern.1em{\raise.47ex\hbox{$\scriptscriptstyle |$}}
             \kern-.40em{\sf C}}
\def\Z{{\sf Z\kern-.45em Z}}

\def\I{{\rm i}}                  
\def\D{{\rm d}}                  

\newcommand{\drond}[2]{\frac{\partial #1}{\partial #2}} 

\def\tL{{\tilde L}}
\def\Dv{\delta\varepsilon}

\begin{document}

\selectlanguage{english}

\title{Individual energy level distributions for one-dimensional 
       diagonal and off-diagonal disorder}

\author{Christophe Texier}

\date{26th June 2000}

\maketitle	

{\small
\noindent
D\'epartement de Physique Th\'eorique, Universit\'e de Gen\`eve, 24~quai 
Ernest Ansermet, CH-1211 Gen\`eve~4, Switzerland.\\
{\bf E-mail:} 
texier@kalymnos.unige.ch
}

\begin{abstract}

We study the distribution of the $n$-th energy level for two
different one-dimensional random potentials. 
This distribution is shown to be related to the distribution of the distance 
between two consecutive nodes of the wave function. 

We first consider the case of a white noise potential and study the 
distributions of energy level both in the positive and the negative
part of the spectrum. It is demonstrated that,
in the limit of a large system ($L\to\infty$), the distribution of the 
$n$-th energy level is given by a scaling law which is shown to be 
related to the extreme value statistics of a set of independent variables.

In the second part we consider the case of a supersymmetric random Hamiltonian
(potential $V(x)=\phi(x)^2+\phi'(x)$). We study first the case of $\phi(x)$
being a white noise with zero mean.
It is in particular shown that the ground state energy, which behaves on 
average like $\exp{-L^{1/3}}$ in agreement with previous work, is not a 
self averaging quantity in the limit $L\to\infty$ as is seen in the case of 
diagonal disorder.
Then we consider the case when $\phi(x)$ has a non zero mean value.

\end{abstract}


\section{Introduction}

One-dimensional disordered systems have been studied in great detail in the 
past and are still a subject of interest. Examples include recent work on 
the failing of the single parameter scaling for localization near the 
band edge or at strong disorder \cite{DeyLisAlt00}, 
the statistical properties of the Wigner time delay, studied in 
several works \cite{JayVijKum89,ComTex97,TexCom99,Tex99,RamKum00}, as well as 
the correlations of the time delay for different energies \cite{TitFyo00}. 
The distribution of another transport time has also been investigated in 
\cite{BolLamFalPriEps99}.
As a last example let us mention that the ac-conductivity was re-examined in 
\cite{Gog00}.

In this article we are interested in the following one-dimensional 
Hamiltonian:
\be\label{hamil}
H=-\frac{\D^2}{\D x^2} + V(x)
\:,\ee
where $V(x)$ is a random potential (we choose units such that $\hbar=2m=1$
for simplicity). The spectral properties of such Hamiltonians were studied
in different works for various kinds of disorder. The case of 
$\delta$-scatterers of random weights and/or at random positions was examined 
by several authors \cite{Sch57,FriLlo60,BycDyk66a} (see also 
\cite{LifGrePas88}). Here we consider two kinds of random potentials. 

({\bf A}) In the first case the potential is a white noise, {\it i.e.} $V(x)$ 
is distributed with the Gaussian weight:
\be\label{measure}
{\cal D}V(x)\,P[V(x)]={\cal D}V(x)\,\exp-\frac{1}{2\sigma}\int\D x\,V(x)^2
\:.\ee
This implies that $\mean{V(x)}=0$ and $\mean{V(x)V(x')}=\sigma\,\delta(x-x')$,
all other cumulants being zero.
The spectral properties of this model were studied by Halperin \cite{Hal65};
later Berezinski\u{i} developed a diagrammatic method adapted to this 
particular model \cite{Ber74} used to study various quantities in a number of 
works.
We can also mention a recent study of the density of states and level 
statistics of this model using soliton calculus \cite{VorVag95}.
This model is equivalent to the $\delta$-scatterer model in a certain regime:
when the density of impurities is very large compared to their weights. 
Moreover it was shown in \cite{AntPasSly81} that results for the Gaussian
disorder are very generic at high energy: they reproduce what is expected
for any random potential provided that the correlation length of this 
disordered potential (which is zero for the Gaussian disorder) 
and the de Broglie wavelength are the smallest length scales of the problem, 
whatever their relative magnitude is.

({\bf B}) The second disordered potential we will consider belongs to 
another class of random potentials, which has attracted the attention of 
several authors: the case of off-diagonal disorder. The random potential reads
\be
V(x)=\phi(x)^2 + \phi'(x)
\ee
where $\phi(x)$ is random. The Hamiltonian, that may be factorized as 
$H=Q^\dagger Q$ where $Q=-\D_x+\phi(x)$, describes one-dimensional 
supersymmetric quantum mechanics \cite{Wit81}.
A discrete version of this model would be a 1D tight-binding Hamiltonian with
random hopping whereas the discrete version of model {\bf A} is a 1D 
tight-binding Hamiltonian with random site energies.
The interest for Anderson model with off-diagonal disorder was recently 
renewed due to its connection with disordered spin chain models 
\cite{Zim82,FabMel97,SteFabGog98}.
It is also worth mentioning the relation of the supersymmetric Hamiltonian
with the problem of classical diffusion in random media (see for example
\cite{BouComGeoLed90} and references therein); many properties were obtained
in this context very recently using a real space renormalization group
method \cite{LedMonFis99}.
In the case of off-diagonal disorder, the density of states presents different 
kinds of behaviour compared to diagonal disorder, like Dyson singularities 
at band edge \cite{Dys53} (if $\mean{\phi(x)}=0$), or power law singularities 
(if $\mean{\phi(x)}\neq0$)
\cite{OvcEri77,BouComGeoLed87,BouComGeoLed90,ComDesMon95,Mon95,BalFis97,ComTex98,Boc99,Boc00}.
Very recently the density of states for coupled chains with off-diagonal 
disorder was studied \cite{BroMudFur00} showing interesting effect of the 
parity of the number of chains.
For a complete review on one-dimensional disorder systems, the interested 
reader is refered to \cite{LifGrePas88,Luc92}.
In the following we will consider the case where the random function involved
in the potential is distributed according to a Gaussian weight: 
${\cal D}\phi(x)\,P[\phi(x)]
={\cal D}\phi(x)\,\exp-\frac{1}{2g}\int\D x\,[\phi(x)-\mu g]^2$. 

\vspace{0.25cm}

The purpose of this article is to study the distribution of the energy $E_n$ 
of the $n$-th excited state of the Hamiltonian (\ref{hamil}) 
($n+1$-th energy level) considered
on a finite interval of length $L$ with the wave function $\varphi(x)$ 
satisfying Dirichlet boundary conditions: $\varphi(0)=\varphi(L)=0$. 
Let us denote this distribution:
\be
W_n(E) = \mean{ \delta(E-E_n) }
\:,\ee
where the brackets $\mean{\cdots}$ mean average over different configurations 
of the disordered potential $V(x)$ (with respect to measure (\ref{measure}) 
for model {\bf A}). These distributions are related to the average density 
of states per unit length by the relation:
\be\label{DoS}
\rho(E) = \frac1L \sum_{n=0}^\infty W_n(E)
\:.\ee

This problem was studied by Grenkova {\it et al.} \cite{GreMolSud83}
who derived these distributions for the $\delta$-impurity model of Frisch and
Lloyd when the weights of impurities are very large compared to the average 
density of impurity and the energy. We stress that this is not the limit 
where this model is equivalent to the model with the white noise potential 
considered here.

The problem was later addressed by McKean \cite{McK94} who derived the 
distribution for the ground state energy ($n=0$) for $E<0$; he considered 
different boundary conditions for the model originally studied by Halperin
for a potential being a white noise.
We will provide in section \ref{dd} a generalization of the result of McKean 
by giving the distribution for all eigenvalues in both regions of the 
spectrum $E<0$ and $E>0$, provided $|E|\gg\sigma^{2/3}$. We will demonstrate 
that the distribution is given by a scaling law in the limit 
$L\to\infty$ and will show how the parameters scale with the sample size $L$.
This scaling law is similar to the distribution of the extreme value of a set 
of statistically independent random variables \cite{Gum58}. This is in 
agreement with the fact that the eigenvalues are not expected to present 
level repulsion in the limit for which the states are localized, as 
demonstrated by Mol\v{c}anov \cite{Mol81}.

After having considered the case {\bf A} of diagonal disorder we will study
in section \ref{offdd} the problem for off-diagonal disorder {\bf B} at band 
edge and at high energy as well. We will consider first the case $\mu=0$
for which we will study the ground state energy distribution. The analysis
for model {\bf A} cannot be applied in this case and we will need to use 
specific approximations.
Note already that our result for the distribution of the ground state energy 
gives a mean value in agreement with the prediction of Monthus {\it et al.} 
\cite{MonOshComBur96} who showed that the averaged ground state energy 
behaves like $\EXP{-L^{1/3}}$ if the system size is very large, by finding 
a lower bound and an upper bound. 
Moreover, our result shows that the ground state energy is not a self averaging
quantity as for model {\bf A} at low energy, and its distribution presents
a large tail. We also give the distribution $W_n(E)$ in the high energy 
limit.
We finally study the distributions for $\mu\neq0$ in the low energy limit.


\section{Idea of the method}

In this section we give the main ideas of the method we use to derive the 
eigenvalue distribution. We concentrate on model {\bf A} since the
ideas are the same for model {\bf B} apart from small subtleties 
which we will discuss later.

Let us consider $\psi(x;E)$, the solution of the Schr\"odinger equation
$H\psi(x;E) = E\psi(x;E)$ with the boundary condition $\psi(0;E)=0$. 
The boundary condition $\psi(L;E)=0$ is fulfilled whenever the energy $E$ 
coincides with an eigenvalue $E_n$ of the Hamiltonian. In this case,
the wave function 
$\varphi_n(x)=\psi(x;E_n)/\left[\int_0^L\D{x'}\,\psi(x';E_n)^2\right]^{1/2}$ 
has $n$ nodes in the interval $]0,L[$, and two nodes at the boundaries. 
Let us denote by $\ell_m$ the $n+1$ lengths between the nodes.
We consider the Ricatti variable
\be
z(x;E) = \frac{\D}{\D x} \ln|\psi(x;E)|
\:,\ee
which obeys the following equation:
\be\label{Riceq}
\frac{\D}{\D x} z = -E - z^2 + V(x)
\:,\ee
for an initial condition $z(0;E)=+\infty$.
This equation may be viewed as a Langevin equation for a particle located at
$z$ submitted to a force deriving from the unbounded potential 
\be
U(z)=Ez+\frac{z^3}{3}
\ee 
and to a random ``force'' $V(x)$ (white noise). 
Each node of the wave function corresponds to $|z(x)|=\infty$. At ``time''
$x=0$ the ``particle'' starts from $z(0)=+\infty$ and eventually ends at
$z(\ell_1-0^+)=-\infty$ after a ``time'' $\ell_1$. Just after the first node
it then starts again from  $z(\ell_1+0^+)=+\infty$, due to the continuity of 
the wave function. It follows from this picture that the distance 
$\ell_m$ between two consecutive nodes may be viewed as the ``time'' needed 
by the particle to go through the interval $]-\infty,+\infty[$ (the 
``particle'' is emitted from $z=+\infty$ at initial ``time'' and absorbed
when is reaches $z=-\infty$).
Following appendix \ref{escapetime} we introduce the $n$-th moment 
${\cal L}_n(z)$ of the ``time'' ${\cal L}$ the particle takes to reach 
$-\infty$ starting from $z$: 
\be
{\cal L}_n(z) = \mean{ {\cal L}^n \:|\: z(0)=z;\ z({\cal L})=-\infty }
\:.\ee
According to (\ref{recumom}) or (\ref{recumom2}) these moments satisfy the 
following recursion relations:
\bea\label{recuLn}
{\cal L}_n(z) &=& \frac{2 n}{\sigma}
\int_{-\infty}^z\D z'\,\EXP{\frac2\sigma U(z')}
\int_{z'}^{+\infty}\D z''\, \EXP{-\frac2\sigma U(z'')} \, {\cal L}_{n-1}(z'')
\:,\\
{\cal L}_0(z) &=& 1
\:.\eea
The boundary condition at $a=+\infty$ is chosen to be a reflecting one. This
choice, which simplifies calculations, is not important since the particle 
can never go back to this edge of the interval and necessarily ends at 
$b=-\infty$.

The moments of the lengths $\ell_m$ between consecutive nodes of the wave 
function are:
\be
\mean{ \ell^n } = {\cal L}_n(+\infty)
\:.\ee
It is worth mentioning that the $n+1$ lengths are statistically independent
because each time the variable $z$ reaches $-\infty$, it loses the memory of 
its earlier history since it is brought back to the same initial condition
and $V(x)$ is $\delta$-correlated. This remark is a crucial point for the 
derivation of $W_n(E)$.

However the fact that $V(x)$ is $\delta$-correlated is not essential to ensure 
the statistical independence of the $\ell_m$'s. Indeed, imagine that 
correlations of $V(x)$ are short range, on a scale $x_c$. Equation 
(\ref{Riceq}) shows that when $z$ starts from $+\infty$ at initial ``time'',
it needs a ``time'' $\Delta x$ to reach the region in $z$-space where $U(z)$
is of order or smaller than $\sigma$ and where the presence of the random 
``force'' $V(x)$ matters for the evolution of $z$. The ``time'' $\Delta x$,
during which the dynamics of $z$ is governed only by the deterministic 
``force'' $-U'(z)$, can be defined as 
$\Delta x=\int_{z_0}^\infty\frac{\D z}{E+z^2}$ where $U(z_0)=\sigma$;
we have $\Delta x\sim1/k$ for $|E|\gg\sigma^{2/3}$,
and $\Delta x\sim\sigma^{-1/3}$ for $|E|\ll\sigma^{2/3}$.
If $z$ follows the deterministic evolution from $+\infty$ during a ``time''
$\Delta x$ sufficient for $V(x)$ to decorrelate, then the $\ell_m$'s are 
statistically independent; this occurs when $x_c$ is much smaller than the 
smallest length scale among $1/k$ and $\sigma^{-1/3}$.

Let us finally write the equation satisfied by the generating function 
of the moments of the traversal ``time'' ${\cal L}$
\be\label{genfunh}
h(\alpha,z) = \mean{ \EXP{-\alpha{\cal L}} \:|\: z(0)=z;\:z({\cal L})=-\infty }
\:.\ee
According to (\ref{eqgenfu}) it obeys:
\be\label{eqpourh}
G_z \, h(\alpha,z) = \alpha \, h(\alpha,z)
\:,\ee
where
\be\label{bfpegen}
G_z = -U'(z) \partial_z + \frac{\sigma}{2} \partial_z^2
\ee
is the generator of the backward Fokker Planck equation (BFPE) associated with 
the stochastic differential equation (\ref{Riceq}). 
$-U'(z)=-\partial_zU(z)$ is the force deriving from the potential.
The generating function satisfies the boundary conditions:
\bea
\partial_z h(\alpha,z)|_{z=+\infty} &=& 0 \:,\\
h(\alpha,-\infty) &=& 1
\:.\eea

Coming back to the initial problem, our goal is to compute the probability
for the energy of the $n$-th excited state to be $E$. This occurs if the 
sum of the $n+1$ distances between the nodes is equal to the length of the 
system: $L=\sum_{m=1}^{n+1}\ell_m$. As it was stated above the $\ell_m$'s are 
independent variables and 
$\proba\left[L=\sum_{m=1}^{n+1}\ell_m\right]$ is given in terms of the 
distribution of the variables $\ell_m$. 
For the different cases we will analyze throughout this article, we will 
initially examine the distribution $P(\ell)$, enabling us to find $W_n(E)$.


\section{Diagonal disorder: white noise potential \label{dd}}

\subsection{Distribution of the distance between consecutive nodes of the wave 
            function}

We study the distribution $P(\ell)$ of the distance $\ell$ between two 
consecutive nodes of the wave function. Let us first note that the average
integrated
density of states per unit length $N(E)=\int_{-\infty}^E\D E'\,\rho(E')$
gives the number of states below energy $E$ per unit length, which is also 
the number of nodes of the wave function of energy $E$ per unit length, or
in other words the inverse of the average distance between two 
consecutive nodes:
\be\label{meanl}
\mean{\ell} = {\cal L}_1(+\infty) = N(E)^{-1}
\:.\ee
The average integrated density of states is given by calculating 
(\ref{recuLn}) for $n=1$ in which it is possible to perform integration
over $z''$ \cite{Hal65,LifGrePas88}:
\bea\label{idoshalp}
N(E)&=&\frac{(\sigma/2)^{1/3}}{\sqrt\pi}
     \left[ \int_0^\infty\frac{\D{y}}{\sqrt{y}}\:
            \EXP{-(\frac{y^3}{12}+\frac{E}{(\sigma/2)^{2/3}}y)} \right]^{-1} \\
    &=&\frac{(\sigma/2)^{1/3}}{\pi^2}
     \left[  {\rm Ai}^2\left(\frac{-E}{(\sigma/2)^{2/3}}\right)
           + {\rm Bi}^2\left(\frac{-E}{(\sigma/2)^{2/3}}\right) \right]^{-1}
\:;\eea
${\rm Ai}(x)$ and ${\rm Bi}(x)$ are Airy functions. This result, 
given by Halperin \cite{Hal65}, was 
first mentioned in \cite{FriLlo60} as an approximation for the integrated 
density of states for a potential consisting of randomly dropped 
$\delta$-scatterers in the limit of high density of scatterers.

\begin{figure}[!ht]
\begin{center}
\includegraphics[scale=1]{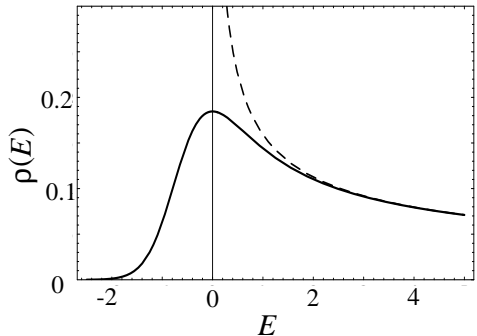}
\end{center}
\caption{Density of states \protect\cite{Hal65} per unit length for 
         $\sigma=1$, given by (\ref{idoshalp}). 
         Dashed: free density of states.}
\end{figure}

We now examine the limit of high energy $|E|/\sigma^{2/3}\to\infty$ in the 
negative and in the positive part of the spectrum.


\mathversion{bold}
\subsubsection{Negative part of the spectrum $E=-k^2$: trapping of the 
               Ricatti variable}
\mathversion{normal}

\begin{figure}[!ht]
\begin{center}
\includegraphics[scale=1]{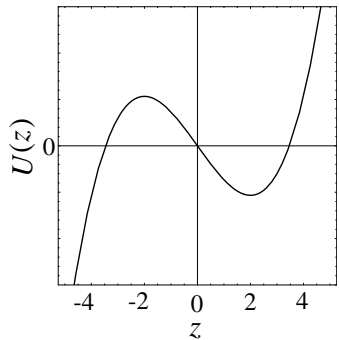}
\end{center}
\caption{The potential that traps the Ricatti variable $z$ for $\sigma=1$ and 
         $E=-4$.\label{pot2}}
\end{figure}

If $E<0$ the potential $U(z)$ possesses a local minimum at $z=\sqrt{-E}=k$ 
that may trap the Ricatti variable (see figure \ref{pot2}). 
In the limit $|E|/\sigma^{2/3}\gg1$, the 
well is very deep (or the diffusion very small) and when traveling from 
$+\infty$ to $-\infty$, the Ricatti variable spends most of the ``time'' in 
the well. Then we expect that the average ``time'' $\mean{\ell}$ is given 
by the Arrhenius law and that its distribution is a Poisson distribution,
as demonstrated in appendix \ref{escapetime}.
Expanding the potential $U(z)$ in the neighbourhood of its two local
extrema
\bea
&& U(z) \APPROX{z\sim k} 
       -\frac{2k^{3}}{3} + k \, (z-k)^2 \\
&& U(z) \APPROX{z\sim-k} 
        \frac{2k^{3}}{3} - k \, (z+k)^2 
\:,\eea
equation (\ref{arrhenlaw}) gives 
\be\label{IDoSapprox}
\mean{\ell}={\cal L}_1(+\infty) = N(E)^{-1}
\simeq \frac{\pi}{\sqrt{-E}} \exp\frac{8}{3\sigma}(-E)^{3/2}
\:.\ee
This argument was first used by Jona-Lasinio \cite{Jon83} to find the 
exponential factor of the integrated density of states. We stress 
here that it also gives the correct pre-factor, that may be checked by 
extracting the limiting behaviour of (\ref{idoshalp}).

As we have shown in appendix \ref{escapetime} the distribution of the time 
spent in the well is a Poisson law:
\be\label{dol-}
P(\ell) = N(E)\,\exp{-\ell\,N(E)}
\:.\ee
This equation will be the starting point to find the distribution $W_n(E)$
in section \ref{disten}.


\mathversion{bold}
\subsubsection{Positive part of the spectrum $E=+k^2$: 
               small disorder expansion}
\mathversion{normal}

\begin{figure}[!ht]
\begin{center}
\includegraphics[scale=1]{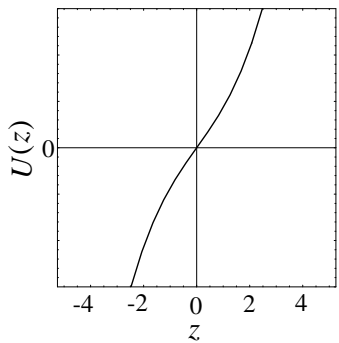}
\end{center}
\caption{Potential for $\sigma=1$ and $E=+4$.\label{pot1}}
\end{figure}

In the positive part of the spectrum, the potential in which the Ricatti 
variable evolves has no local extremum (see figure \ref{pot1}) and therefore 
the length $\mean{\ell}$ does not anymore follow an Arrhenius law.

In the absence of diffusion ($\sigma=0$) the ``time'' needed by the Ricatti 
variable to go from $+\infty$ to $-\infty$ is $\ell = \pi/k$, which may
be found either by integrating (\ref{Riceq}) in the absence of the potential
$V(x)$ or by taking the limit $\sigma\to0$ in (\ref{meanl},\ref{idoshalp}).
We can expect that, for sufficiently weak disorder, the length is weakly 
fluctuating around its mean value. We are now going to show that the 
distribution of $\ell$ is a narrow Gaussian distribution in this limit.
For this purpose we will analyze the moments 
$\smean{\ell^n}={\cal L}_n(+\infty)$, which
is more conveniently achieved by studying a generating function. Instead
of the generating function of the moments (\ref{genfunh}), it is more 
advantageous to consider the generating function of the cumulants:
\be\label{genfunw}
w(\alpha,z) = \ln h(\alpha,z)
\:.\ee
It follows from (\ref{eqpourh}) and (\ref{bfpegen}) that
\be\label{eqpourw}
-(k^2+z^2)\partial_z w + \frac{\sigma}{2} 
\left[ \partial_z^2 w + (\partial_z w)^2\right] = \alpha
\:.\ee
The solution may be obtained through a perturbative expansion in the 
parameter $\sigma$. We write
\be
w = w^{(0)} + w^{(1)} + w^{(2)} + \cdots
\:,\ee
where $w^{(n)}=O(\sigma^n)$. The boundary condition for $w$ is:
$w(\alpha,z\to-\infty)=0$. The $0$-th order of (\ref{eqpourw}) gives
\be\label{w0}
w^{(0)}(\alpha,z) = -\alpha \int_{-\infty}^z\D z'\frac{1}{k^2+z^2}
=-\frac{\alpha}{k}\left( \arctan\frac{z}{k} +\frac{\pi}{2} \right)
\:.\ee
The integral is the ``time'' needed to go from $z$ to $-\infty$ in the absence
of diffusion as it is clear from (\ref{Riceq}).
The $n$-th order of (\ref{eqpourw}) gives:
\be
w^{(n)}(\alpha,z) = \frac{\sigma}{2} \int_{-\infty}^z \frac{\D z'}{k^2+z'^2}
\left[
  \partial_{z'}^2 w^{(n-1)}(\alpha,z') + 
  \sum_{m=0}^{n-1} \partial_{z'}w^{(m)}(\alpha,z')
                 \:\partial_{z'}w^{(n-1-m)}(\alpha,z')
\right]
\:.\ee
In particular we have:
\be\label{w1}
w^{(1)}(\alpha,z) = \frac{\sigma}{2k^3}
\left[
  - \frac{\alpha}{2k}\frac{1}{(1+z^2/k^2)^2}
  + \left(\frac{\alpha}{k}\right)^2 \int_{-\infty}^{z/k}\frac{\D x}{(1+x^2)^3}
\right]
\:,\ee
\be
w^{(2)}(\alpha,z) = \left(\frac{\sigma}{2k^3}\right)^2
\left[
    \frac{\alpha}{k}\int_{-\infty}^{z/k}\D x\frac{2-10x^2}{(1+x^2)^5}
  + \left(\frac{\alpha}{k}\right)^2  \frac{5}{4(1+z^2/k^2)^4}
  - \left(\frac{\alpha}{k}\right)^3 \int_{-\infty}^{z/k}
                                    \frac{2\D x}{(1+x^2)^5}
\right]
\:.\ee
Equation (\ref{w1}) gives the dominant contribution to the second cumulant
$\cum{\ell^2}$ of the length. Explicit computation of terms of $w$ of 
higher order in $\sigma$ is not required and we only need to know their 
behaviour with $\alpha$. Using a recursion argument it is possible to show 
that $w^{(n)}$ is a polynomial of degree $n+1$ in $\alpha$:
\be\label{wn}
w^{(n)}(\alpha,z) = \sigma^n \sum_{m=1}^{n+1} \alpha^m \, c_m(z)
\:.\ee
We now extract from these expressions the information about the cumulants
of $\ell$:
\be
w(\alpha,+\infty) = \sum_{m=1}^\infty \frac{(-\alpha)^m}{m!} \cum{\ell^m}
\:.\ee
Using (\ref{w0},\ref{w1}) we see that:
\bea
w(\alpha,+\infty)&=& 
w^{(0)}(\alpha,+\infty) + w^{(1)}(\alpha,+\infty) + O(\sigma^2) \\
&=& - \alpha\frac{\pi}{k} 
    + \frac{\alpha^2}{2!} \left(\frac{\pi}{k}\right)^2 \frac{3\sigma}{8\pi k^3}
    + O(\sigma^2)
\:;\eea
then
\bea
\mean{\ell}  &=& \frac{\pi}{k} + O(\sigma^2) \label{cum1}\\
\cum{\ell^2} &=& \left(\frac{\pi}{k}\right)^2 \frac{3\sigma}{8\pi k^3} 
                  + O(\sigma^2) \label{cum2}
\:.\eea
The $n$-th cumulant is given by the term proportional to $\alpha^n$ in 
$w(\alpha,\infty)$.
Equation (\ref{wn}) shows that the term of lowest order in $\sigma$ containing 
$\alpha^n$ is $w^{(n-1)}$, then
\be
\cum{\ell^n}  =  O(\sigma^{n-1})
\:.\ee
The $n$-th order cumulant for $n>2$ characterizes fluctuations that are 
negligible, in the high energy limit, compared to the fluctuations described 
by the second cumulant:
\be
\frac{\cum{\ell^n}}{\cum{\ell^2}^{n/2}} 
= O\left[ \left(\frac{\sigma}{k^3}\right)^{\frac{n}{2} - 1}\right]
{ {\raisebox{-.3cm}{$\longrightarrow$}} \atop 
  {\scriptstyle{\sigma/k^3\to0}} } 0
\:.\ee
Since the second cumulant is dominating it follows that the distribution 
of $\ell$ is Gaussian in this limit:
\be\label{dol+}
P(\ell) = \frac{1}{\sqrt{2\pi\cum{\ell^2}}}
\exp-\frac{\left(\ell-\mean{\ell}\right)^2}{2\cum{\ell^2}}
\:.\ee
For the positive part of the spectrum, (\ref{dol+}) shows that the 
fluctuations of $\ell$ are small compared to its average value,
in contrast to what happens in the negative part of the spectrum
where (\ref{dol-}) shows that the fluctuations of $\ell$ are of the 
same order as the average value.


\subsubsection*{The structure of the wave function}

Since we are dealing with one-dimensional disordered system, the wave 
functions are expected to be localized, {\it i.e.} decreasing with an 
exponential damping on a length scale being by definition the localization 
length $\lambda$. Let us recall that $\lambda\simeq\frac{8E}{\sigma}$ for 
$E\to+\infty$ and $\lambda\simeq\frac{1}{\sqrt{-E}}$ for $E\to-\infty$
(see \cite{AntPasSly81} for example).
The results we have derived for the distribution $P(\ell)$ show that in
the limit $E\to+\infty$, the consecutive nodes of the wave function are 
separated by weakly fluctuating distances of order ${\pi}/{k}$, which is 
much smaller than $\lambda$.
In contrast, in the limit $E\to-\infty$ the distance $\ell$ is distributed 
according to a Poisson law (\ref{dol-}) which means that it is probable to 
find two consecutive nodes as close as possible. However the typical scale 
of the length is much larger than the localization length:
$\ell\sim N(E)^{-1}\simeq\lambda\pi\exp\frac{8}{3\sigma}(-E)^{3/2}\gg\lambda$.


\subsection{Distribution of individual energy level \label{disten}}

We now derive the distribution $W_n(E)$. As we have mentioned above, $E$ 
coincides with the eigenvalue $E_n$ if the sum of the $n+1$ lengths between 
nodes is equal to the length $L$.


\mathversion{bold}
\subsubsection{Low energy $E=-k^2$}
\mathversion{normal}

We follow here the idea McKean used to find the ground state energy 
distribution \cite{McK94}. 
The probability that the energy $E$ is between two consecutive energies
is:
\bea\label{theproba}
&&\proba\left[E_{n-1}<E<E_{n}\right]
= \proba\left[\ell_1+\cdots+\ell_n < L <\ell_1+\cdots+\ell_n+\ell_{n+1}
          \right] \\
&&= \int_0^\infty\D\ell_1\,P(\ell_1)\cdots
    \int_0^\infty\D\ell_{n+1}\,P(\ell_{n+1})\,
    \heav\left( \ell_1+\cdots+\ell_{n+1} - L \right)\,
    \heav\left( L- \ell_1-\cdots-\ell_{n}    \right)
\,,\:\eea
where $\heav(x)$ is the Heaviside function. Using (\ref{dol-}) we get:
\be\label{proba1}
\proba\left[E_{n-1}<E<E_{n}\right]=\frac{(L\,N(E))^n}{n!} \, \EXP{-L\,N(E)} 
\:.\ee
Introducing the joint distribution for the eigenvalues
\be
W_{0,1,\cdots,n,\cdots}(X_0,X_1,\cdots,X_n,\cdots)
=\mean{ \prod_{m=0}^\infty\delta(X_m-E_m) }
\ee
and differentiating (\ref{theproba}) with respect to $E$ gives
\bea
&&\frac{\D}{\D E} \proba\left[E_{n-1}<E<E_{n}\right] \nonumber\\
&&=\frac{\D}{\D E} 
\int\D X_0\,\D X_1\cdots\D X_n\cdots
W_{0,1,\cdots,n,\cdots}(X_0,X_1,\cdots,X_n,\cdots)\,
\heav(E-X_{n-1})\,\heav(X_n-E) \hspace{0.5cm}\label{inutile}
\:.\eea
Differentiation of the integrand of (\ref{inutile})
gives two terms. In each term the Heaviside function $\heav(X_n-X_{n-1})$
does not play any role since the joint probability for the different
energies is proportional to the following product of Heaviside 
functions:
\be
W_{0,1,\cdots,n,\cdots}(X_0,X_1,\cdots,X_n,\cdots) \propto
\heav(X_1-X_0)\,\heav(X_2-X_1)\cdots\heav(X_n-X_{n-1})\cdots
\:.\ee
Then we find:
\be\label{dproba1}
\frac{\D}{\D E} \proba\left[E_{n-1}<E<E_{n}\right]= W_{n-1}(E) - W_n(E)
\:.\ee
For $n=0$ we recover the result of McKean \cite{McK94} for the ground 
state energy distribution:
\be
\int_E^\infty\D E'\,W_0(E') = \EXP{-L\,N(E)} 
\:.\ee
Using (\ref{proba1}) and (\ref{dproba1}), it is now easy to show that:
\be\label{TheResult}
W_n(E) = L\rho(E) \, \frac{(L\,N(E))^n}{n!} \, \EXP{-L\,N(E)} 
\:.\ee
This result has a clear meaning since $L\rho(E)$ gives the probability 
to find an energy at $E$, the factor $\frac{x^n}{n!}\EXP{-x}$ ``compelling''
the number of states below $E$, $x=LN(E)$, to be close to $n$.

Let us remark that the relation (\ref{DoS}) is satisfied.

We now analyze this result in more detail and demonstrate that 
the distribution (\ref{TheResult}) has a scaling form in the limit
$L\to\infty$. We use the approximated expression (\ref{IDoSapprox}) of the 
integrated density of states per unit length to
write the distribution $W_n(E)$ as:
\be
W_n(E) = \frac{2^{n+3}(n+1)^{n+1}}{n!} \tL^{n+1} \EXP{-g(E)}
\ee
with
\be
g(E)   = (n+1)\frac{8(-E)^{3/2}}{3} 
         + 2(n+1)\tL\sqrt{-E}\,\EXP{-\frac83(-E)^{3/2}} - \frac{n+2}{2}\ln(-E)
\:,\ee
where we have introduced: 
\be\label{Ltilde}
\tL=\frac{L\,\sigma^{1/3}}{2\pi(n+1)}
\:.\ee
We set $\sigma=1$ for simplicity since it will be easy to recover its 
dependence at the end (the dimension of $\sigma$ is: 
$[\sigma]={\rm length}^{-3}={\rm energy}^{3/2}$).

We first look for the typical value of the $n$-th excited state energy
$E_n^{\typ}$. It is the solution of the equation $g'(E)=0$. The variable
$Y={8}(-E_n^{\typ})^{3/2}$ is the solution of 
\be
\tL = \frac{ Y-\frac{n+2}{n+1} }{ Y-1 }\,\frac{\EXP{Y/3}}{Y^{1/3}}
\:.\ee
In the limit $L\to\infty$ we find 
$Y=3\ln(\tL) + \ln(3\ln(\tL)) + O\left[\frac{\ln(3\ln(\tL))}{\ln(\tL)}\right]$,
that is:
\be\label{entyp}
E_n^{\typ}(L)=-\left(\frac{3\sigma}{8}\ln\tL\right)^{2/3}\times
\left[
  1 + \frac29 \frac{\ln(3\ln\tL)}{\ln\tL} 
    + O\left(\frac{\ln^2(3\ln\tL)}{\ln^2\tL}\right)
\right]
\:,\ee
where we keep the first correction to the dominant term in $\ln^{2/3}\tL$
because it is larger than the difference between the typical values 
of two consecutive levels (\ref{dentyp}).
This behaviour was given as a good approximation of $E_0$ by McKean 
\cite{McK94}. Note that the $n$-dependence enters only in $\tL$.

We will need the maximum value of the distribution, which is:
\be
W_n(E_n^{\typ})=\frac{2(n+1)^{n+1}\EXP{-(n+1)}}{n!}
\left(3\ln\tL\right)^{1/3}\times
\left[
  1 + O\left(\frac{\ln(3\ln\tL)}{\ln\tL}\right)
\right]
\:.\ee
We now study the derivatives of $g(E)$ at $E=E_n^{\typ}$. By definition the 
first derivative vanishes:
\be\label{dg1}
g'(E_n^{\typ})=O\left(\frac{\ln(3\ln\tL)}{\ln^{2/3}\tL}\right)
\:.\ee 
After a little of algebra, we find for the $m$-th derivative:
\be\label{dgm}
g^{(m)}(E_n^{\typ}) = 2^m(n+1)
\left(3\ln\tL\right)^{m/3}\times
\left[
  1 + O\left(\frac{\ln(3\ln\tL)}{\ln\tL}\right)
\right],
\hspace{1cm} m\geq2
\:.\ee
The second derivative defines the scale of the fluctuations of $E_n$:
\be\label{fluctu}
\delta E_n = \frac{1}{\sqrt{g^{(2)}(E_n^{\typ})}}
=\sigma^{2/3}\frac{ \left(3\ln\tL\right)^{-1/3} }{2\sqrt{n+1}}
\times
\left[
  1 + O\left(\frac{\ln(3\ln\tL)}{\ln\tL}\right)
\right]
\:.\ee
Note that the width of $W_n(E)$ decreases as $L\to\infty$. The relative 
fluctuations tends to zero, that is $E_n$ is self averaging in this limit.

Moreover it is possible to introduce the function $\omega_n(X)$:
\be
W_n(E) = \frac{1}{\delta E_n} \ 
         \omega_n\left(\frac{E-E_n^{\typ}}{\delta E_n}\right)
\:,\ee
with
\be
\omega_n(X)=\frac{W_n(E_n^{\typ})}{\sqrt{g^{(2)}(E_n^{\typ})}}
\exp-\sum_{m=1}^\infty \frac{a_m}{m!}X^m
\:,\ee
where $a_m=g^{(m)}(E_n^{\typ})/[g^{(2)}(E_n^{\typ})]^{m/2}$.
Using (\ref{dg1},\ref{dgm}) it is easy to see that $a_1=0$ and 
$a_m=(n+1)^{1-m/2}$ in the limit $L\to\infty$. Then we find for the 
scaling function:
\be\label{TheResultbis}
  \omega_n(X)=\frac{(n+1)^{n+\frac12}}{n!}
  \exp{\left(\sqrt{n+1}\:X -(n+1)\,\EXP{{X}/{\sqrt{n+1}}}\right)}
\:.\ee
A similar result was
obtained by Grenkova {\it et al.} \cite{GreMolSud83} for the $\delta$-impurity
model of Frisch and Lloyd, in the limit of low impurity density which
has no equivalent in the model we consider here. These authors found the same
scaling distribution (\ref{TheResultbis}) whereas the scaling 
(\ref{entyp},\ref{fluctu}) is different (it is of course model-dependent).

It is interesting to note that (\ref{TheResultbis}) is related to extreme
value statistics. If we consider a set $\{x_k\}$ of ${\cal N}\to\infty$ 
statistically 
independent variables distributed according to the same law and order them:
$x_0<x_1<x_2<\cdots$, then the distribution of $x_n$ has the form of 
(\ref{TheResultbis}) up to a rescaling. $\omega_0(X)$ is the distribution of 
the most negative, etc. This problem was studied by E.~Gumbel in 1935. 
For more details about extreme value statistics, see \cite{Gum54,Gum58}
(see appendix \ref{extreme}). The distribution (\ref{TheResultbis}) is the 
extreme value distribution for any distribution of the so called exponential 
type (unlimited domain and finite moments).
Moreover the form of the scaling (\ref{entyp},\ref{fluctu}) allows us to 
find the tail of the distribution $p(E)$ of one of those variables:
$p(E)\sim\exp-\frac83(-E)^{3/2}$. This tail is not surprising since it is
precisely the behaviour of the density of states (see also the discussion
in the conclusion). 
The remark that the eigenvalues are distributed as a set of statistically 
independent variables is in agreement with the expected absence of level 
repulsion for one-dimensional disordered systems in the localized regime 
\cite{Mol81,GreMolSud83}.
Equation (\ref{TheResultbis}) allows us to compute the generating function
\be
{\cal G}_n(k)=\int_{-\infty}^{+\infty}\D X\,\omega_n(X)\,\EXP{kX} 
= \frac{\Gamma(n+1+k\sqrt{n+1})}{\Gamma(n+1)}\:(n+1)^{-k\sqrt{n+1}}
\:.\ee
The expansion of the logarithm of the generating function in powers of $k$
gives the cumulants of the distribution (\ref{TheResultbis}):
\bea
\mean{X}  &=& \sqrt{n+1}\,\big[\psi(n+1)-\ln (n+1)\big] \\
\cum{X^m} &=& (n+1)^{m/2}\,\psi^{(m-1)}(n+1) \hspace{1cm} {\rm for}\ m\geq2
\:,\eea
where $\psi(z)=\frac{\D}{\D z}\ln\Gamma(z)$ is the digamma function.
It is possible to show that (\ref{TheResultbis}) converges to a Gaussian 
distribution in the large $n$ limit. Using the Stirling formula we find
$\ln{\cal G}_n(k)\simeq-\frac{k}{2\sqrt{n}}+\frac{k^2}{2}$, {\it i.e.}
\be
\lim_{n\to\infty} \omega_n(X) = \frac1{\sqrt{2\pi}}\EXP{-\frac{X^2}{2}}
\:.\ee

\begin{figure}[!ht]
\begin{center}
\includegraphics[scale=0.9]{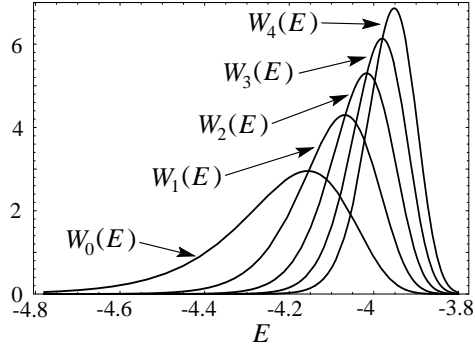}
\caption{The distributions for the first five energy levels. $\sigma=1$
         and $L=10^{10}$.\label{fourdist}}
\end{center}
\end{figure}

It is also possible to compute the mean level spacing $\mean{\Delta_n}$ between
the $n$-th excited state and the $n+1$-th one. Equation (\ref{entyp}) shows 
that:
\be\label{dentyp}
E_{n+1}^{\typ} - E_{n}^{\typ}
= \sigma^{2/3}\frac12\ln\frac{n+2}{n+1}\:\left(3\ln\tL\right)^{-1/3}
\:,\ee
where $\tL$ is still given by (\ref{Ltilde}) for the $n$-th excited state 
(and not the length associated with the $n+1$-th one).
Since the mean energy is $\mean{E_n}=E_{n}^{\typ}+\delta E_{n}\mean{X}$, we 
eventually find:
\be\label{mls}
\mean{\Delta_n} =\mean{E_{n+1}-E_n}=
\frac{\sigma^{2/3}}{2(n+1)}\left(3\ln\tL\right)^{-1/3}
\:.\ee
This result shows that the average distance between two consecutive levels
is of the same order as the fluctuations (\ref{fluctu}) of the position 
of those levels.
In other words, the distributions $W_n(E)$ are overlapping functions, as 
represented in figure \ref{fourdist}.
Now, if we consider that $L$ is fixed, we can see from (\ref{fluctu}) and 
(\ref{mls}) that, apart from the unimportant $n$-dependence of $\tL$ in the 
logarithm, the mean level spacing decreases like 
$\mean{\Delta_n}\propto1/n$ whereas the fluctuations decreases like 
$\delta E_n\propto1/\sqrt{n}$ for large $n$. As $n$ increases for fixed $L$, 
the consecutive distributions $W_n(E)$ become more and more overlapped, 
this is also suggested in figure \ref{fourdist}.

\mathversion{bold}
\subsubsection{High energy $E=+k^2$}
\mathversion{normal}

In order to find the distribution $W_n(E)$ we use the same idea as before:
the probability that $E=E_n$ is related to the probability that the sum of 
the  $n+1$ lengths between the nodes of the wave function is $L$.
Let us introduce $\Lambda=\sum_{m=1}^{n+1}\ell_m$. We have
$W_n(E)\D E={\cal P}_n(\Lambda)\D\Lambda$, where ${\cal P}_n(\Lambda)$ is the 
distribution function for $\Lambda$. Since the $\ell_m$'s are statistically
independent ${\cal P}_n(\Lambda)$ is easily found from (\ref{dol-}) or 
(\ref{dol+}) (note that ${\cal P}_0(\Lambda)=P(\ell=\Lambda)$).
If $E=E_n$, the sum of the lengths between nodes coincides with $L$: 
$\Lambda=L$. In this case we have $\Lambda\,N(E)=n+1$. Differentiating
this latter equation we get $N(E)\D\Lambda+\Lambda\,\rho(E)\D E=0$, that is
$\frac{\D\Lambda}{\D E}=-\frac{\Lambda\,\rho(E)}{N(E)}$. It follows that:
\be\label{relPW}
W_n(E) = \frac{L\,\rho(E)}{N(E)} \, {\cal P}_n(\Lambda=L)
\:.\ee
Using (\ref{dol-}), this equation allows us to get straightforwardly 
(\ref{TheResult}). In the positive part of the spectrum, since the 
distribution (\ref{dol+}) is Gaussian, the distribution ${\cal P}_n(\Lambda)$
is also Gaussian and we get:
\be
W_n(E) = \frac{ L\,\rho(E) }{ \sqrt{2\pi(n+1)N(E)^2\cum{\ell^2}} }
\exp{ -\frac{ (L\,N(E)-n-1)^2 }{ 2N(E)^2\cum{\ell^2} } }
\:,\ee
with $\cum{\ell^2}$ being given by (\ref{cum2}).

We now go to the variable $k=\sqrt{E}$ for clarity. We write:
\be\label{Wn+prov}
W_n(E)\D E = \D k\frac{1}{\sqrt{2\pi s(k)}}
\exp{ -\frac{(k-(n+1)\frac{\pi}{L})^2}{2s(k)} }
\:,\ee
with $s(k)=(n+1)\frac{3\pi}{8}\frac{\sigma}{k^3L^2}$. The maximum value 
of the distribution corresponds to $k=k^0_n=(n+1)\frac{\pi}{L}$, which is
the de Broglie wavelength in the absence of the disordered potential. 
We remember that we are dealing with a high energy limit 
$E\sim\frac{(n+1)^2}{L^2}\gg\sigma^{2/3}$, which may be conveniently
written as:
\be\label{validity}
\frac{1}{n+1}\frac{L}{\lambda_n}\ll1
\:,\ee 
where we have introduced the localization length $\lambda_n$ associated with 
an energy $E_n^0=(k_n^0)^2=(n+1)^2\frac{\pi^2}{L^2}$. We recall that 
$\lambda=\frac{8 E}{\sigma}$ for $E\gg\sigma^{2/3}$ 
\cite{AntPasSly81}. The condition (\ref{validity}) is fulfilled either
for a delocalized regime $L\ll\lambda$, or for a localized regime 
$L\gg\lambda$ provided that $n$ is sufficiently large.
Note that there is no restriction on $L$ for the derivation of 
$W_n(E)$.

Due to the condition of validity (\ref{validity}),
it is possible to neglect the dependence of $s(k)$ on $k$ and replace $s(k)$
by $s(k_0^n)$; then the distribution $W_n(E)$ is a Gaussian distribution 
\be\label{Wn+}
W_n(E) = \frac1{\sqrt{2\pi\delta E_n^2}}
\EXP{-\frac{(E-\mean{E_n})^2}{2\delta E_n^2}}
\ee
of mean:
\be\label{Enmoy+}
\mean{E_n} = (n+1)^2\frac{\pi^2}{L^2}
\ee
and width
\be\label{fluctu+}
\delta E_n = \sqrt{\frac{3\sigma}{2L}}
\:.\ee
According to (\ref{validity}) the relative fluctuations are necessarily small:
$\frac{\delta E_n^2}{\mean{E_n}^2}=\frac{3}{\pi^2(n+1)^2}
\frac{L}{\lambda_n}\ll1$.
More interesting is to compare these fluctuations to the mean level spacing 
$\mean{\Delta_n}=\mean{E_{n+1}-E_n}\simeq 2(n+1)\frac{\pi^2}{L^2}$. Then
\be\label{LRornot}
\frac{\delta E_n^2}{\mean{\Delta_n}^2} 
\simeq \frac{3}{\pi^2}\frac{L}{\lambda_n}
\:.\ee
This condition tells us that the fluctuations $\delta E_n$ of the position 
of $E_n$ are larger than the mean level spacing $\mean{\Delta_n}$ if we are 
in a localized regime; the distributions $W_n(E)$ are overlapping functions. 
This agrees with the fact that no level repulsion is expected in this regime 
\cite{Mol81,GreMolSud83} where the level spacing is believed to be 
distributed according to a Poisson law \cite{Mol81}.
In the delocalized regime $L\ll\lambda$, the fact that the distributions 
$W_n(E)$ are non-overlapping indicate level repulsion.

It is worth mentioning that (\ref{Enmoy+},\ref{fluctu+}) may be found by a 
simpler, although less systematic, perturbative argument: the energy of the
$n+1$-th level is, up to first order pertubation theory
$E_n\simeq E_n^0+\bra{\psi_n^0} V(x)\ket{\psi_n^0}$, with 
$\psi_n^0(x)=\sqrt{\frac{2}{L}}\sin k_n^0x$. Then it is straightforward to see
that $\smean{E_n}$ is given by (\ref{Enmoy+}) and that 
$\delta E_n^2=\smean{[\int_0^L\D x\,V(x)\psi_n^0(x)^2]^2}
=\sigma\int_0^L\D x\,\psi_n^0(x)^4$ leads to (\ref{fluctu+}).

\vspace{0.5cm}

Let us end the section with a remark concerning the ground state energy.
We have noticed that the ground state energy is a self averaging quantity
($\mean{E_0}\gg\delta E_0$). Its behaviour with $L$ is 
$E_0\sim-\ln^{2/3}L$ for $L\to\infty$, then it is vanishing 
$E_0\sim0$ for a size $L\sim\sigma^{-1/3}$ and behaving like 
$E_0\sim1/L^2$ for $L\to0$.


\section{Off-diagonal disorder: supersymmetric random Hamiltonian 
         \label{offdd}}

The Hamiltonian we consider in this section is the following supersymmetric
Hamiltonian:
\be\label{Hsusy}
H_S=-\frac{\D^2}{\D x^2} + \phi(x)^2 + \phi'(x)
\:.\ee
The spectral and localization properties of this Hamiltonian were studied 
for various kinds of disorder. The case of white noise was analyzed in
\cite{OvcEri77,BouComGeoLed87,BouComGeoLed90}. The case of a random 
telegraph process was also studied in detail \cite{ComDesMon95,Mon95}. 
Note that the Hamiltonian (\ref{Hsusy}) is the square of a Dirac Hamiltonian 
with a random mass. The spectral properties of 1D random Dirac Hamiltonians 
were studied in a more general situation by Bocquet \cite{Boc99,Boc00}.
His analysis was extended very recently to study the distribution 
function of the local density of states \cite{BunMcK00}; these authors 
used replica trick and supersymmetry and obtained for the supersymmetric 
Hamiltonian in a high energy limit a result similar to the one derived by 
Altshuler and Prigodin \cite{AltPri89} for model {\bf A} using the 
Berezinski\u{i} technique.

As it was mentioned in the introduction, we focus on the situation where
$\phi(x)$ is a white noise:
\be\label{meassusy}
{\cal D}\phi(x)\,P[\phi(x)]
={\cal D}\phi(x)\,\exp-\frac{1}{2g}\int\D x\,\left[\phi(x)-\mu g\right]^2
\:.\ee
We recall that the integrated density of states is in this case 
\cite{OvcEri77,BouComGeoLed87,BouComGeoLed90,LifGrePas88}:
\be\label{IDoSsusy}
N(E)=\frac{2g}{\pi^2}\frac{1}{J_\mu^2(\sqrt{E}/g)+N_\mu^2(\sqrt{E}/g)}
\:,\ee 
where $J_\nu(z)$ and $N_\nu(z)$ are the Bessel functions of first and second 
kind, respectively.

We now consider the situation where 
\mathversion{bold}$\mu=0$\mathversion{normal} and will discuss the case 
$\mu\neq0$ in a last section.
In general, the Hamiltonian (\ref{Hsusy}) may possess a zero mode, however, 
for the problem of interest here, it can not satisfy the Dirichlet 
boundary conditions; we say that the supersymmetry is broken.
Due to the supersymmetric structure of the Hamiltonian the spectrum of $H_S$
is positive.
It is possible to rewrite the Schr\"odinger equation 
$H_S\varphi(x)=k^2\varphi(x)$ as two coupled first order differential 
equations:
\bea
Q^\dagger \chi(x) &=& k \varphi(x) \\
Q \varphi(x)      &=& k \chi(x)
\:,\eea
where $Q=-\D_x+\phi(x)$ and $Q^\dagger=\D_x+\phi(x)$.
As in the previous section, the first step is to study the statistics of the
distances between the nodes of the wave function. For the supersymmetric
Hamiltonian it is not very convenient to consider the ``Ricatti'' variable
$z=\frac{\chi}{\varphi}$ because it is not an additive process but a 
multiplicative process: $\D_xz=k+kz^2-2\phi(x)z$.
To help the discussion we introduce two intermediate variables, a phase 
variable $\vartheta(x)$ and an envelope function $\exp\xi(x)$:
\bea
\varphi(x) &=&  \EXP{\xi(x)} \sin\vartheta(x) \\
\chi(x)    &=& -\EXP{\xi(x)} \cos\vartheta(x)
\:.\eea
These two functions obey the set of coupled stochastic differential equations,
written in the Stratonovich convention:
\bea \label{eqtheta}
\frac{\D}{\D x}\vartheta &=& k + \phi(x)\sin2\vartheta 
\hspace{1cm} \mbox{(Stratonovich)} \\
\frac{\D}{\D x}\xi       &=&   - \phi(x)\cos2\vartheta
\hspace{1.35cm} \mbox{(Stratonovich)}
\:.\eea
The fact that the periodicity in $\vartheta$ of these two equations is 
$\pi/2$, and that $V(x)$ is $\delta$-correlated, means that the distance 
separating the two points where 
$\vartheta=m\pi/2$ and $\vartheta=(m+1)\pi/2$ depends only on $V(x)$ between
these two points  (the periodicity in $\vartheta$ is $\pi/2$ and not $\pi$ 
because the sign can always be absorbed in the function $\phi(x)$ if it is 
a white noise of zero mean). 
Then the statistically independent random variables to be considered in a 
first step are not the distances $\ell$ between successive nodes but rather 
the distances, denoted $\Lambda$, separating the points where $\vartheta(x)$ 
takes a value equal to a multiple integer of $\pi/2$ (the point where 
$(\vartheta(x)\:{\rm mod}\:\pi)=\pi/2$ corresponds to a local extremum of 
the oscillatory part of the wave function).
The length $\ell$ between two consecutive nodes of the wave function is then 
the sum of two independent $\Lambda$'s.

\begin{figure}[!ht]
\begin{center}
\includegraphics[scale=1]{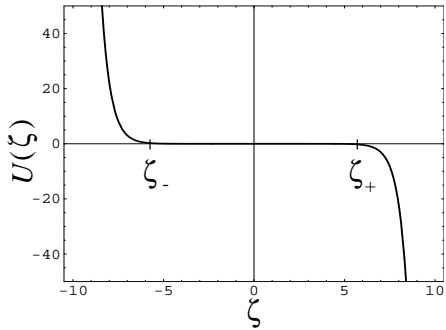}
\caption{Potential ``seen'' by the variable $\zeta$ for $k=10^{-5}$.
         \label{potzeta}}
\end{center}
\end{figure}

It is more convenient to introduce a process which is additive in the noise
$\phi(x)$. It is easy to see that this is achieved by the change of variable
$\zeta(x)=\frac12\ln|\tan\vartheta(x)|$. This variable obeys the stochastic
differential equation:
\be \label{eqzeta}
\frac{\D}{\D x}\zeta = k\cosh2\zeta + \phi(x)
\:.\ee
This is a Langevin equation for a ``particle'' of position $\zeta$ traveling 
from $-\infty$ to $+\infty$ in a potential 
\be
U(\zeta)=-\frac{k}{2}\sinh2\zeta
\ee 
(see figure \ref{potzeta})
and feeling the random ``force'' $\phi(x)$. $\Lambda$ is the ``time'' the 
``particle'' needs to go through the interval. As for the 
model {\bf A}, we are going to study the statistical properties of the 
distances $\ell$ between the nodes, sum of two independent variables $\Lambda$.
The knowledge of the distribution of $\ell$ will allow us to find the 
distribution of the energies.


\subsection{Distribution of distances between the nodes}

We introduce the moments of the ``time'' needed by the variable $\zeta(x)$ to 
reach $+\infty$ starting from the position $\zeta$:
\be
\tilde\Lambda_n(\zeta) = 
\smean{ \tilde\Lambda^n \:|\: \zeta(0)=\zeta;\ \zeta(\tilde\Lambda)=+\infty }
\:.\ee
They are given by (see appendix \ref{escapetime}):
\bea
\tilde\Lambda_n(\zeta) &=& \frac{2 n}{g}
\int_\zeta^{+\infty}\D\zeta'\,\EXP{\frac2g U(\zeta')}
\int_{-\infty}^{\zeta'}\D\zeta''\, 
\EXP{-\frac2g U(\zeta'')} \, \tilde\Lambda_{n-1}(\zeta'')
\\
\tilde\Lambda_0(\zeta) &=& 1
\:.\eea
The moments of the random variable of interest are 
$\mean{\Lambda^n}=\tilde\Lambda_n(-\infty)$.


\mathversion{bold}
\subsubsection{Low energy limit: $E\ll g^2$}
\mathversion{normal}

At low energy $k\ll g$, the potential is still a
monotonic function and there is no process of trapping of the variable
$\zeta$ by a well as for the low energy limit of model {\bf A}, which means 
that we have to develop a specific approximation scheme. For this purpose we 
start by studying the average time $\tilde\Lambda_1(\zeta)$ to go from
$\zeta$ to $+\infty$:
\be\label{temps1}
\tilde\Lambda_1(\zeta) = \frac{2}{g}
\int_\zeta^{+\infty}\D\zeta'\,\EXP{-\frac{k}{g}\sinh2\zeta'}
\int_{-\infty}^{\zeta'}\D\zeta''\, 
\EXP{\frac{k}{g}\sinh2\zeta''}
\:.\ee
We introduce $\zeta_\pm$, the two solutions of the equation 
$\frac{\D^2}{\D\zeta^2}\exp\left(\frac{k}{g}\sinh2\zeta\right)=0$. In the low 
energy limit: $\zeta_\pm\simeq\pm\frac12\ln\frac{g}{k}$.
The positions $\zeta_\pm$ are the crossover points where the force deriving
from the potential $U(\zeta)$ is of the same order as the random ``force''
$\phi(x)$.
The study of (\ref{temps1}) leads us to distinguish three regions to which
the initial condition can belong to:

\noindent ({\it i}) $\zeta_+<\zeta$
\be
\tilde\Lambda_1(\zeta) \simeq \frac{1}{g}\EXP{-2(\zeta-\zeta_+)}
\:.\ee

\noindent ({\it ii}) $\zeta_-<\zeta<\zeta_+$
\be
\tilde\Lambda_1(\zeta) \simeq 
\frac{1}{g}\left[\left(\zeta_+-\zeta_-\right)^2 
                 - \left(\zeta-\zeta_-\right)^2 \right]
+ \frac{1}{g} 
\:.\ee

\noindent ({\it iii}) $\zeta<\zeta_-$
\be
\tilde\Lambda_1(\zeta) \simeq 
\frac{1}{g} \left[1-\EXP{2(\zeta-\zeta_-)}\right]
+ \frac{1}{g}\left(\zeta_+-\zeta_-\right)^2 + \frac{1}{g} 
\:.\ee

\noindent
What can we learn from these behaviours~? 
({\it i}) if $\zeta$ starts from the neighbourhood of $\zeta_+$, it needs a 
``time'' of order $1/g$ to reach $+\infty$.
({\it ii}) When $\zeta$ starts somewhere in $[\zeta_-,\zeta_+]$ the ``time'' 
needed to reach $+\infty$ is dominated by the first term, which behaves like 
$\zeta^2$, {\it i.e.} like the ``time'' for a free diffusive ``particle''.
({\it iii}) When $\zeta$ starts from $-\infty$ it needs a ``time'' $1/g$ to 
reach $\zeta_-$; then $\zeta$ travels from $\zeta_-$ to $\zeta_+$ in a 
``time'' 
$\frac{({\rm distance})^2}{\rm diffusion}
=\frac1g(\zeta_+-\zeta_-)^2\simeq\frac1g\ln^2\frac{g}{k}$ and 
eventually ends at $+\infty$ after an additional ``time'' $1/g$.

The physical picture for the motion of the fictitious particle of position 
$\zeta$ we get from these results is now very clear. $\zeta$ travels from 
$-\infty$ to $\zeta_-$ very quickly due to the potential only; 
for $\zeta\in[\zeta_-,\zeta_+]$ the potential becomes negligible compared to 
the random force which is of order $g$ and $\zeta$ evolves due to the random
force only. It increases like the position a free diffusive particle until 
it reaches $\zeta_+$ from where it goes very rapidly to $+\infty$. 

We are now going to use this picture to study the distribution of $\Lambda$.
We are interested in the characteristic function for the traveling time 
\be
h(\alpha,\zeta) = 
\smean{ \EXP{-\alpha\tilde\Lambda} \:|\: 
        \zeta(0)=\zeta;\ \zeta(\tilde\Lambda)=+\infty }
\:.\ee
From the Langevin equation (\ref{eqzeta}) we get the BFPE generator and we 
see that $h$ obeys (see appendix \ref{escapetime})
\be\label{eqhex}
\left(k\cosh2\zeta\:\partial_\zeta + \frac{g}{2}\partial_\zeta^2\right)
h(\alpha,\zeta) = \alpha \: h(\alpha,\zeta)
\ee
with boundary conditions:
\bea\label{bounhex1}
\partial_\zeta h(\alpha,\zeta)\big|_{-\infty} &=& 0 \:,\\ \label{bounhex2}
h(\alpha,+\infty) &=& 1
\:.\eea
Since the traveling ``time'' is dominated by the ``time'' spent in the 
region $[\zeta_-,\zeta_+]$ where the diffusion is free, it means that 
the characteristic function in the limit $k\ll g$ is approximatively 
given by the solution of the equation for the free diffusion on this
finite interval. Then equations (\ref{eqhex},\ref{bounhex1},\ref{bounhex2})
may be replaced by the following equation
\be
\frac{g}{2}\partial_\zeta^2 h(\alpha,\zeta) = \alpha \: h(\alpha,\zeta)
\ee
with boundary conditions:
\bea
\partial_\zeta h(\alpha,\zeta_-) &=& 0 \:,\\
h(\alpha,\zeta_+) &=& 1
\:.\eea
The first condition is a reflection condition since the variable $\zeta$
``sees'' a steep wall in $\zeta_-$. The second condition is an absorption 
condition since as $\zeta$ reaches $\zeta_+$, it eventually ends to $+\infty$
after a negligible time. Since we are dealing with free diffusion, the 
solution is then very easy to find:
\be
h(\alpha,\zeta) = 
\frac{\cosh\left[\sqrt{\frac{2\alpha}{g}}(\zeta-\zeta_-)\right]}
     {\cosh\left[\sqrt{\frac{2\alpha}{g}}(\zeta_+-\zeta_-)\right]}
\:.\ee
The characteristic function for the length $\Lambda$ is:
\be
\smean{\EXP{-\alpha\Lambda}} = h(\alpha,\zeta_-) 
= \frac{1}{\cosh\sqrt{\alpha B}}
\:,\ee
where we have introduced 
$B=\frac2g\ln^2\frac{g}{k}=\frac{1}{2g}\ln^2\frac{g^2}{E}$. Since we are 
in fact interested in the distance $\ell$ between the nodes of the wave 
function, which is given as the sum of two independent $\Lambda$'s, we give 
the characteristic function for $\ell$:
\be\label{dlsusy0}
\smean{\EXP{-\alpha\ell}} = \frac{1}{\cosh^2\sqrt{\alpha B}}
\:.\ee
The inverse Laplace transform gives the distribution function:
\be\label{InvLap}
P(\ell) = \frac1B \int_{-\I\infty}^{+\I\infty}\frac{\D q}{2\I\pi}
\frac{\EXP{q\ell/B}}{\cosh^2\sqrt{q}}
\:,\ee
where the integral is taken over a Bromwitch contour. $\cosh\sqrt{q}$ is an
analytic function in the variable $q$, whose zeros are: 
$q_m=-\frac{\pi^2}{4}(2m+1)^2$ for $m\in\NN$ (note that there is no branch
cut since $\cosh$ is an even function). We can expand $\cosh\sqrt{q}$ in the
neighbourhood of its zeros and we get:
\be\label{cosq}
\cosh\sqrt{q} \APPROX{q\sim q_m}\frac{(-1)^m}{\pi(2m+1)}(q-q_m)
\left[1+\frac{1}{\pi^2(2m+1)^2}(q-q_m)+\cdots\right]
\:.\ee
We use the residue's theorem to evaluate integral (\ref{InvLap}); we can 
indeed check that the contribution of integrals over the large semi-circles 
needed to close the contour in the complex plane are vanishing in the limit 
of their radius going to infinity. Then we get:
\be\label{dol0susy}
P(\ell) = \theta(\ell) \frac1B \sum_{m=0}^\infty
\left[\frac{\ell}{B}\pi^2(2m+1)^2 - 2\right]
\EXP{-\frac{\pi^2}{4}(2m+1)^2\ell/B} 
\:.\ee
Using the identities \cite{gragra}: 
$\sum_{m=0}^\infty\frac{1}{(2m+1)^2}=\frac{\pi^2}{8}$,
$\sum_{m=0}^\infty\frac{1}{(2m+1)^4}=\frac{\pi^4}{96}$ and 
$\sum_{m=0}^\infty\frac{1}{(2m+1)^6}=\frac{\pi^6}{960}$, we can check the 
normalization and get the two first cumulants:
\bea
\mean{\ell} &=& B = \frac{1}{2g}\ln^2\frac{g^2}{E} \:,\\
\cum{\ell^2} &=& \frac13\mean{\ell}^2
\:.\eea
The average distance between consecutive nodes is the inverse of the 
integrated density of states. We can check from the exact result
(\ref{IDoSsusy}) that the limit behaviour of $N(E)$ is indeed in perfect
agreement with the result we find here
$\mean{\ell}=N(E)^{-1}\simeq\frac{1}{2g}\ln^2\frac{g^2}{E}$, which gives
us a certain confidence in the approximation we have made to calculate the
characteristic function $h(\alpha,\zeta)$.
Since $B=\mean{\ell}$ it can be replaced by $N(E)^{-1}$ and we can rewrite 
the distribution as
\be\label{dol0susy2}
P(\ell) = N(E) \, \varpi_0(N(E) \ell)
\:,\ee
where
\be\label{pi0susy}
\varpi_0(X) =  \theta(X) \sum_{m=0}^\infty
\left[\pi^2(2m+1)^2X - 2\right] \EXP{ -\frac{\pi^2}{4}(2m+1)^2 X } 
\ee
replaces the exponential function obtained for the low energy limit in model 
{\bf A} (\ref{dol-}).

We can extract the limiting behaviours of the distribution (\ref{dol0susy}).
For this purpose we introduce the $\theta$-function (not to be confused with
the Heaviside function $\heav(x)$)
\be\label{eq1}
\tilde\theta(x)=\sum_{m=0}^\infty\EXP{-(2m+1)^2x}
=\frac14\sqrt{\frac{\pi}{x}}
\left(1 + 2\sum_{m=1}^\infty(-1)^m\EXP{-m^2\frac{\pi^2}{4x}}\right)
\:,\ee
(this is the elliptic theta function $\vartheta_1(\pi/2\,|\,4\I x/\pi)$  
\cite[8.180]{gragra}). 
The distribution (\ref{pi0susy}) is related to the $\theta$-function by:
\be\label{eq2}
\varpi_0(X) = -2 \left( 2X\,\drond{}{X} + 1 \right)
\tilde\theta\left(\frac{\pi^2}{4}X\right)
\:.\ee
Using (\ref{eq1},\ref{eq2}) we can find an expression adapted for the
limit $X\to0$:
\be\label{Plim1}
\varpi_0(X) = \frac{4}{\sqrt{\pi}}\:\frac{\heav(X)}{X^{3/2}} 
\sum_{m=1}^\infty(-1)^{m+1} m^2\EXP{-m^2/X}
\APPROX{X\to0} \frac{4}{\sqrt{\pi}}\:\frac{\heav(X)}{X^{3/2}}\:\EXP{-1/X}
\:.\ee
The tail of the distribution is exponential and dominated by the term $m=0$
in (\ref{pi0susy}):
\be\label{Plim2}
\varpi_0(X) \APPROX{X\to\infty} \pi^2 X \EXP{ -\frac{\pi^2}{4} X }
\:.\ee

\begin{figure}[!ht]
\begin{center}
\includegraphics{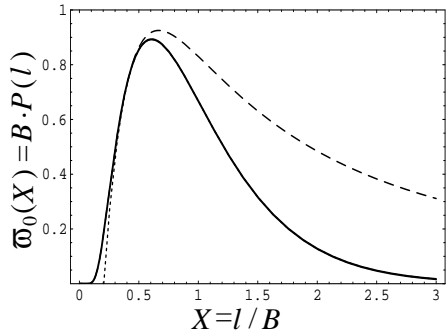}
\end{center}
\caption{Function $\varpi_0(X)$ giving the  distribution of the distance 
between consecutive nodes of the wave function for the supersymmetric 
Hamiltonian in the limit $E\to0$.
Dashed: term $m=1$ of (\ref{Plim1}); 
Dotted: term $m=0$ of (\ref{pi0susy}).\label{Pellsusy}}
\end{figure}

\subsubsection*{Structure of the wave function at low energy}

It is well known that the supersymmetric Hamiltonian (\ref{Hsusy}) exhibits
a delocalization transition at zero energy \cite{BouComGeoLed90}. As energy 
goes to zero, the localization length behaves like 
$\lambda\simeq\frac1g\ln(4g^2/E)$. The distribution we have just found for the 
distances between the nodes shows that two consecutive nodes are separated 
by a distance of order $B=\frac1g\ln^2(g^2/E)$ which is much larger than the 
localization length $\lambda$. It is also worth mentioning that this 
distance is the correlation length appearing in the average Green's function 
\cite{BouComGeoLed90}.
By contrast to what happens for the model {\bf A} at low energy, where the
nodes of the wave function can be arbitrarily close, the nodes of the 
wave function for the supersymmetric model {\bf B} with $\mu=0$ are 
extremely unlikely to be closer than a distance of order 
$\frac1g\ln^2(g^2/E)$: the behaviour of the distribution $P(\ell)$ (figure 
\ref{Pellsusy}) indicates a ``repulsion'' of the nodes of the wave function.


\mathversion{bold}
\subsubsection{High energy limit ($E\gg g^2$): small disorder expansion}
\mathversion{normal}

At high energy we perform the same perturbative analysis in the disordered
strength $g$ as for model {\bf A}, hence we do not enter into the details. 
The generating function for the cumulants 
$w(\alpha,\zeta)=\ln h(\alpha,\zeta)$ obeys:
\be
k\cosh2\zeta\:\partial_\zeta w + \frac{g}{2} 
\left[ \partial_\zeta^2 w + (\partial_\zeta w)^2\right] = \alpha
\:.\ee
We solve this equation perturbatively in $g$:
$w = w^{(0)} + w^{(1)} + \cdots$, where $w^{(n)}=O(g^n)$. 
Since the reflection condition is at $\zeta=-\infty$ and the absorbing one 
at $\zeta=+\infty$, we have now: $w(\alpha,+\infty)=0$ to ensure 
$h(\alpha,+\infty)=1$.
To zeroth order we find:
\be
w^{(0)}(\alpha,\zeta) = -\frac{\alpha}{k} 
\int_\zeta^{+\infty}\D\zeta'\frac{1}{\cosh2\zeta'}
\ee
and to order $n$:
\be
w^{(n)}(\alpha,\zeta) = \frac{g}{2k} \int_\zeta^{\infty}
\frac{\D\zeta'}{\cosh2\zeta'}
\left[
  \partial_{\zeta'}^2 w^{(n-1)}(\alpha,\zeta') + 
  \sum_{m=0}^{n-1} \partial_{\zeta'}w^{(m)}(\alpha,\zeta')
                 \:\partial_{\zeta'}w^{(n-1-m)}(\alpha,\zeta')
\right]
\:.\ee
The information of interest is contained in the first two orders:
\be
w(\alpha,-\infty)= 
- \alpha\frac{\pi}{2k} + \frac{\alpha^2}{2!}\frac{\pi g}{4k^3} + O(g^2)
\:,\ee
that give the cumulants of the variable $\Lambda$. Let us recall that the
length $\ell$ is the sum of two independent $\Lambda$'s, then:
\bea
\mean{\ell}  &=& \frac{\pi}{k} + O(g^2) \:,\\
\cum{\ell^2} &=& \left(\frac{\pi}{k}\right)^2 \frac{g}{2\pi k} + O(g^2) 
\:.\eea

Since the cumulants of higher orders are small, $\cum{\ell^n}=O(g^{n-1})$,
the distribution of $\ell$ is Gaussian in the small disorder limit as for
the high energy limit of model {\bf A}, given by (\ref{dol+}).


\subsection{Distribution of energy level}


\mathversion{bold}
\subsubsection{Low energy: $E\ll g^2$}
\mathversion{normal}

\begin{figure}[!ht]
\begin{center}
\includegraphics{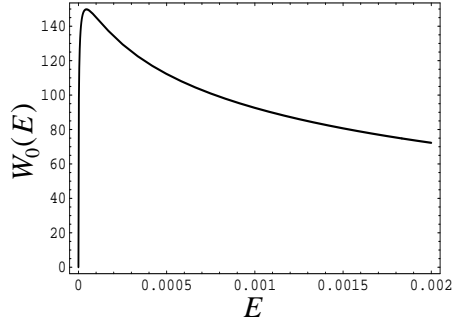}
\end{center}
\caption{Distribution $W_0(E)$ for the supersymmetric potential with $\mu=0$.
         $L=10$, $g=1$.\label{figW0susy}}
\end{figure}

We first consider the distribution for the ground state energy and will give
an explicit form for the excited state energy distribution without going 
further into the calculation to avoid technical increasing complexity. 
The probability for the ground state to be at a given energy is proportional 
to the probability that the distance between the two first nodes of the 
solution of the Schr\"odinger equation is equal to the length of the 
disordered region. According to (\ref{relPW}) we have:
\be
W_0(E) = \frac{L\rho(E)}{N(E)} P(\ell=L)
\:.\ee
Using (\ref{dol0susy2}), we see that this distribution is a scaling function 
of the variable:
\be\label{scalsusy0}
X=L\,N(E)=\frac{2 g L}{\ln^2(g^2/E)}
\:,\ee
which is the averaged number of states below $E$ for a system of length $L$.
Then 
\be\label{W0susy}
W_0(E) = L\rho(E)\:\varpi_0(L\,N(E))
\:.\ee 
From (\ref{Plim1},\ref{Plim2}) we see that 
the distribution $W_0(E)$ presents a log-normal behaviour at low energy 
\be \label{W0lim1}
W_0(E) \simeq \frac{8}{\sqrt{2\pi gL}} \frac1E 
\exp-\frac{\ln^2(g^2/E)}{2gL} \hspace{0.5cm} 
{\rm for} \hspace{0.5cm} E\ll g^2\EXP{-\sqrt{2gL}}
\ee
and the following behaviour at large $E$ (however smaller than $g^2$, not to
be out of the range of validity of the approximation we have made):
\be \label{W0lim2}
W_0(E) \simeq \frac{8\pi^2g^2L^2}{E\ln^5(g^2/E)}
\exp-\frac{\pi^2gL}{2\ln^2(g^2/E)} \hspace{0.5cm} 
{\rm for} \hspace{0.5cm} g^2\EXP{-\sqrt{2gL}}\ll E\ll g^2
\:,\ee
(see figure \ref{figW0susy}).

We may now proceed to study several aspects of the distribution $W_0(E)$.
The typical value of the distribution is given by the limiting behaviour 
of the distribution for small $E$ (\ref{W0lim1}):
\be
E_0^{\typ} \simeq g^2 \EXP{-g L}
\:.\ee
It is also interesting to estimate the median value:
$\int_0^{E_0^{\rm med}}\D E\ W_0(E)=1/2$. Assuming that we can also consider 
the limit behaviour (\ref{W0lim1}), we find that it is the solution of 
$\Phi\left(\frac{\ln(g^2/E_0^{\rm med})}{\sqrt{2gL}}\right)\simeq7/8$, where
$\Phi(z)$ is the error function. Then:
\be
E_0^{\rm med} \sim g^2 \EXP{-c\sqrt{gL}}
\ee
($c$ is a numerical factor)
which is at the boundary of the domain where approximation (\ref{W0lim1}) 
holds.

We may also compute the moments of the ground state energy:
\be\label{momprov}
\smean{E_0^\beta} = \int_0^\infty \D E\,E^\beta\,W_0(E)
=\int_0^\infty \D X\, \varpi_0(X)\,\EXP{-\beta\sqrt{2gL/X}}
\:.\ee
If $\beta>0$ the exponential select the tail (\ref{Plim2}) of the distribution:
\be
\smean{E_0^\beta} \simeq \pi^2 \int_0^\infty \D X\, X
\,\EXP{-\frac{\pi^2}{4}X-\beta\sqrt{\frac{2gL}{X}}}
\:.\ee
This integral is easily worked out by the steepest descent method, and we 
eventually find:
\be\label{momsusy}
\smean{E_0^\beta} \APPROX{L\to\infty}
\frac{16\beta g^{2\beta}}{\sqrt{6\pi}} \, (gL)^{1/2} \, 
\exp-\frac32(\pi^2\beta^2 gL)^{1/3}
\hspace{1cm} {\rm for} \ \beta>0
\:.\ee
In particular, the mean value of the ground state energy behaves like 
$\mean{E_0}\sim\EXP{-\tilde{c}\,L^{1/3}}$, which is much larger than the 
median value. 
This result is in agreement with the work of Monthus {\it et al.} 
\cite{MonOshComBur96} who found upper and lower bounds using a perturbative 
expression for the ground state energy as a functional of $\phi(x)$.
The atypical behaviour of the positive moments with $n$ like 
$\smean{{E_0}^n}\sim b^{n^{2/3}}$ may also be recovered with the method used
in \cite{MonOshComBur96} as remarked by Oshanin \cite{Osh00}.
 
It is also possible to compute the negative moments. If $\beta<0$, the 
exponential in (\ref{momprov}) selects the origin of the distribution. The 
steepest descent method gives:
\be
\label{momsusy2}
\smean{E_0^\beta} \APPROX{L\to\infty}
8 g^{2\beta}\exp\frac12\beta^2gL
\hspace{1cm} {\rm for} \ \beta<0
\:,\ee
which presents a dependence in $\beta$ characteristic of the log-normal 
behaviour (\ref{W0lim1}).

We now make some remarks.

\vspace{0.25cm}

\noindent ({\it i})
It is worth mentioning that the fluctuations are behaving at large $L$ 
like $\delta E_0^2\sim\exp-\frac32(4\pi^2 gL)^{1/3}$ and are much larger than
the mean value $\mean{E_0}^2\sim\exp-\frac32(8\pi^2 gL)^{1/3}$. Thus the 
ground state energy is not a self averaging quantity in the $L\to\infty$ 
limit as it was the case for the model {\bf A}.

\vspace{0.25cm}

\noindent ({\it ii}) 
A remark related to the previous one. Since the distribution $W_0(E)$ is 
a scaling function of $L/\ln^2(1/E)$, it becomes more and more peaked near zero
when $L\to\infty$. Moreover it is rather obvious that it can not be written
as a scaling function of $(E-E_1(L))/E_2(L)$ where $E_{1,2}(L)$ would take 
into account the $L$-dependence as it was possible for model {\bf A}. In 
other words, none of the moments (\ref{momsusy}) determines the nature of 
the fluctuations at $L\to\infty$ since 
$\smean{E_0^\beta}^{1/\beta}\sim L^{1/2\beta}\exp-\frac32(\pi^2L/\beta)^{1/3}$
has a non trivial dependence in $\beta$.

\vspace{0.25cm}

\noindent ({\it iii}) 
It is interesting to note that the average Green's function, derived in 
\cite{BouComGeoLed90}, presents the same kind of behaviour as the 
distribution $W_0(E)$:\\
$\mean{\bra{x}\frac{1}{E-H_S+\I0^+}\ket{x'}}
\simeq\sum_m c_m \exp\left(-\frac{\pi^2}{2\ln^2E}(2m+1)^2|x-x'|\right)$ 
(in the limit $E\to0$ with $g=1$).

\vspace{0.25cm}

\noindent ({\it iv}) 
The distribution (\ref{pi0susy}) was obtained in \cite{LedMonFis99} in the 
context of the classical diffusion by using a real space renormalization 
group method. In this case the distribution is interpreted as the 
distribution of the smallest relaxation time.

\vspace{0.25cm}

\noindent ({\it v}) 
The distribution (\ref{pi0susy}) has still another interpretation: it is 
related to the distribution of the span of a Brownian motion (see 
\cite{MonOshComBur96} and references therein).

\vspace{0.25cm}

\noindent ({\it vi}) 
The $n$-th excited state energy distribution is related to the distribution 
${\cal P}_n(\Lambda)$ of the sum of the $n+1$ lengths $\ell$ by (\ref{relPW}) 
and may be studied by the same kind of calculations, starting from: 
\be\label{pinsusy}
W_n(E) = L\rho(E)
\int_{-\I\infty}^{+\I\infty}\frac{\D q}{2\I\pi}
\frac{\EXP{q\,L\,N(E)}}{\cosh^{2(n+1)}\sqrt{q}}
\ee
where we have used the fact in (\ref{dlsusy0}) that 
$B=\mean{\ell}=N(E)^{-1}=\frac{1}{2g}\ln^2(g^2/E)$.
For example we find that the first excited state energy distribution, 
$W_1(E)=L\rho(E)\:\varpi_1(L\,N(E))$, involves the scaling function:
\bea\label{om1susy}
\varpi_1(X) &=&  \heav(X) \sum_{m=0}^\infty
\bigg[\frac{\pi^4}{6}(2m+1)^4 X^3 - 2\pi^2(2m+1)^2 X^2  \nonumber\\
&& \hspace{3cm}
+2\left(\frac{\pi^2}{3}(2m+1)^2+1\right) X- \frac43\bigg]
\EXP{-\frac{\pi^2}{4}(2m+1)^2X} \\ \label{om1susyB}
&=& \frac{8\,\heav(X)}{3\sqrt{\pi}\,X^{3/2}}
\sum_{m=1}^\infty (-1)^m m^2(m^2-1)\EXP{-m^2/X}
\:,\eea
where $X$ is given by (\ref{scalsusy0}). Its limiting behaviours are:
\bea
\varpi_1(X) && \hspace{-0.6cm}\APPROX{X\to0} 
               \frac{32\,\heav(X)}{\sqrt\pi X^{3/2}} \EXP{-4/X} \:,\\ 
            && \hspace{-0.6cm}\APPROX{X\to\infty} 
               \frac{\pi^4}{6} X^3 \EXP{-\frac{\pi^2}{4}X}
\:.\eea

Note however that for large $n$, since the moments of $\ell$ are finite,
we expect a Gaussian distribution for ${\cal P}_n(\Lambda)$ due to the 
central limit theorem.

\vspace{0.25cm}

\noindent ({\it vii}) 
The distribution $W_0(E)$ is not the distribution of the extreme value of 
a set of independent random variables (see appendix \ref{extreme}) as it is 
the case for model {\bf A}. We will come back to this point at the end.


\mathversion{bold}
\subsubsection{High energy: $E\gg g^2$}
\mathversion{normal}

For high energy the analysis is very similar to the one that was done for 
the model {\bf A} since the distribution of lengths $\ell$'s is the same. 
Only the behaviours of several quantities change. The distribution $W_n(E)$ 
is given by (\ref{Wn+prov}) with $s(k)=(n+1)\frac{\pi g}{2kL^2}$. The 
condition to be at high energy leads to the same expression (\ref{validity})
with the only difference that the localization length for the supersymmetric
model at high energy reaches a constant value: $\lambda\simeq\frac2g$. Then
(\ref{Wn+}) still holds with the mean value of the energy still being the 
free result (\ref{Enmoy+}) and the fluctuation being
\be
\delta E_n = (n+1)\sqrt{\frac{2g\pi}{L^3}}
\:.\ee
The fluctuations now depend on $n$, compared to model {\bf A}, and its
$L$-dependence also changes. However they remain small compared to the 
mean value $\mean{E_n}$ due to condition (\ref{validity}). It is also 
interesting to note that the ratio of the fluctuation and the mean level 
spacing has the same form:
\be
\frac{\delta E_n^2}{\mean{\Delta_n}^2} 
\simeq \frac{2}{\pi^2}\frac{L}{\lambda}
\:,\ee
which means that the absence of level repulsion may only be expected for 
the localized regime.


\mathversion{bold}
\subsection{The case $\mu\neq0$ at low energy ($E\ll g^2$)}
\mathversion{normal}

\begin{figure}[!ht]
\begin{center}
\includegraphics[scale=1]{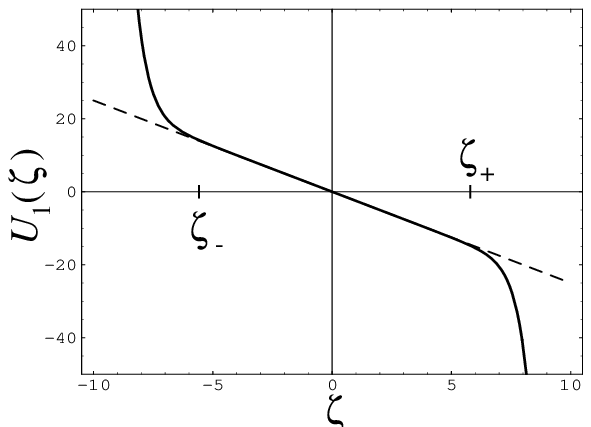}
\hspace{1cm}
\includegraphics[scale=1]{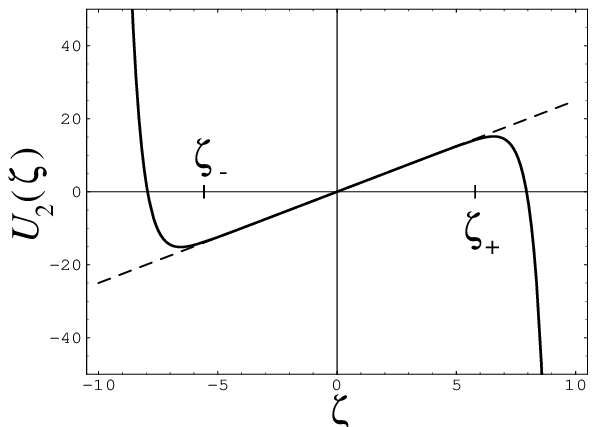}
\end{center}
\caption{The two potentials in which the variable $\zeta$ evolves 
alternatively. \label{potsusy12}}
\end{figure}

We consider in this last section the case where the function $\phi(x)$ is 
distributed according to (\ref{meassusy}) with a 
non zero mean: $\mean{\phi(x)}=\mu g$ (note that $\mu$ is a dimensionless
parameter). We will only discuss the low energy case since we are not 
expecting any modification at high energy.
We have to come back for a while to equation (\ref{eqtheta}) 
for the phase from which we want to get an equation for an additive process.
Performing the change of variable $\zeta=\frac12\ln|\tan\vartheta|$, we have
in fact to distinguish two cases: if 
$(\vartheta\:{\rm mod}\:\pi)\in[0,\pi/2]$ then 
$1/\sin2\vartheta=+\cosh2\zeta$ and we get (\ref{eqzeta}), but if 
$(\vartheta\:{\rm mod}\:\pi)\in[\pi/2,\pi]$ then 
$1/\sin2\vartheta=-\cosh2\zeta$ and we arrive
at $\D_x\zeta=-k\cosh2\zeta+\phi(x)$. The change of the sign in the 
potential is related to the fact that $\zeta$ decreases if $\vartheta$
increases if $(\vartheta\:{\rm mod}\:\pi)\in[\pi/2,\pi]$. So a more convenient 
change of variable is in fact:
\bea
\zeta&=&\frac12\ln|\tan\vartheta| 
\hspace{1.3cm} {\rm if} \ (\vartheta\ {\rm mod}\ \pi)\in[0,\pi/2] \\
\zeta&=&-\frac12\ln|\tan\vartheta| 
\hspace{1cm} {\rm if} \ (\vartheta\ {\rm mod}\ \pi)\in[\pi/2,\pi]
\:.\eea
With this convention, $\zeta$ is always traveling from $-\infty$ to $+\infty$.
Then $\zeta$ obeys 
\be\label{eqzeta1}
\frac{\D}{\D x}\zeta = k\cosh2\zeta + \phi(x)
\hspace{1cm} {\rm if} \ (\vartheta\ {\rm mod}\ \pi)\in[0,\pi/2]
\ee
and 
\be\label{eqzeta2}
\frac{\D}{\D x}\zeta = k\cosh2\zeta - \phi(x)
\hspace{1cm} {\rm if} \ (\vartheta\ {\rm mod}\ \pi)\in[\pi/2,\pi]
\:.\ee
It follows that we have to consider 
alternatively two different equations. We did not mention this point before
because if $\phi(x)$ is a white noise of zero mean, since $\zeta$ decorrelates
when it reaches $+\infty$, the difference in the sign can always be absorbed
in the white noise. We rewrite these two equations in terms of a normalized 
white noise $\eta(x)$ of zero mean: $\mean{\eta(x)}=0$ and 
$\mean{\eta(x)\,\eta(x')}=\delta(x-x')$:
\bea
\label{eqzmu1}
\frac{\D}{\D x}\zeta &=& k\cosh2\zeta +\mu g+\sqrt{g}\,\eta(x) 
\hspace{1cm} {\rm if} \ (\vartheta\ {\rm mod}\ \pi)\in[0,\pi/2] \\
\label{eqzmu2}
\frac{\D}{\D x}\zeta &=& k\cosh2\zeta -\mu g+\sqrt{g}\,\eta(x) 
\hspace{1cm} {\rm if} \ (\vartheta\ {\rm mod}\ \pi)\in[\pi/2,\pi]
\:.\eea
We have to introduce two potentials $U_1(\zeta)$ and $U_2(\zeta)$:
\bea
U_1(\zeta) &=& -\mu g\zeta -\frac{k}{2}\sinh2\zeta \\
U_2(\zeta) &=& +\mu g\zeta -\frac{k}{2}\sinh2\zeta
\:,\eea
plotted in figure \ref{potsusy12}.

We must also introduce two variables $\Lambda_1$ and $\Lambda_2$ that 
give the ``times'' the variable $\zeta$ needs to go from $-\infty$ to 
$+\infty$ in potential $U_1(\zeta)$ and $U_2(\zeta)$, respectively. For a
drift $\mu=0$ those two variables have the same statistical properties as 
it was implicit in the previous section but if $\mu\neq0$, their 
distributions are different. We are now going to study these two 
distributions, denoted $p_1(\Lambda_1)$ and $p_2(\Lambda_2)$. We introduce
the two characteristic functions for the ``times'' $\tilde\Lambda_{1,2}$ 
needed by the random process $\zeta(x)$ obeying (\ref{eqzmu1},\ref{eqzmu2}) 
to reach $+\infty$ starting from $\zeta$:
\be
h_{1,2}(\alpha,\zeta) = 
\smean{ \EXP{-\alpha\tilde\Lambda_{1,2}} \:|\: 
        \zeta(0)=\zeta;\ \zeta(\tilde\Lambda_{1,2})=+\infty }
\:.\ee
We recall that 
$h_{1,2}(\alpha,-\infty)=\int_0^{\infty}\D\Lambda\,p_{1,2}(\Lambda)
\,\EXP{-\alpha\Lambda}$.
These two functions obey (see appendix \ref{escapetime}):
\be
\left( (k\cosh2\zeta\pm\mu g)\:\partial_\zeta 
       + \frac{g}{2}\partial_\zeta^2\right)
h_{1,2}(\alpha,\zeta) = \alpha \: h_{1,2}(\alpha,\zeta) 
\:,\ee
where the upper sign (here $+$) corresponds to $1$ and the lower sign
(here $-$) corresponds to $2$. The boundary conditions are as usual 
$\partial_\zeta h_{1,2}(\alpha,\zeta)\big|_{-\infty}=0$ and 
$h_{1,2}(\alpha,+\infty)=1$. Since we are dealing with the limit $E\to0$, we
can make the same approximation as for the $\mu=0$ case, {\it i.e.} replace
the previous equations by an equation for the free diffusion on the interval
$[\zeta_-,\zeta_+]$, since the time spent out of this interval is negligible.
Then we have to solve:
\be
\left( \pm\mu g\:\partial_\zeta + \frac{g}{2}\partial_\zeta^2\right)
h_{1,2}(\alpha,\zeta) = \alpha \: h_{1,2}(\alpha,\zeta) 
\ee
with the boundary conditions:
\bea
\partial_\zeta h_{1,2}(\alpha,\zeta_-) &=& 0 \:,\\
h_{1,2}(\alpha,\zeta_+) &=& 1
\:.\eea
The solution is easily found:
\be
h_{1,2}(\alpha,\zeta) = \EXP{\mp\mu(\zeta-\zeta_+)}
\frac{ \cosh\gamma(\zeta-\zeta_-) 
       \pm \frac{\mu}{\gamma}\sinh\gamma(\zeta-\zeta_-)}
     {\cosh\gamma(\zeta_+-\zeta_-) 
       \pm \frac{\mu}{\gamma}\sinh\gamma(\zeta_+-\zeta_-)}
\:,\ee
where
\be
\gamma=\sqrt{\frac{2\alpha}{g} + \mu^2}
\:.\ee
The characteristic functions for $\Lambda_{1,2}$ are:
\be
\smean{\EXP{-\alpha\Lambda_{1,2}}} = h_{1,2}(\alpha,\zeta_-) 
=\frac{\EXP{\pm\mu(\zeta_+-\zeta_-)}}
      {\cosh\gamma(\zeta_+-\zeta_-) 
       \pm \frac{\mu}{\gamma}\sinh\gamma(\zeta_+-\zeta_-)}
\:.\ee

We can more conveniently consider the generating function 
$w_{1,2}(\alpha,\zeta_-)=\ln h_{1,2}(\alpha,\zeta_-)$ for the cumulants
of $\Lambda_{1,2}$. In the limit of low energy, when the condition
\be\label{condition}
\mu(\zeta_+-\zeta_-)=\mu\ln g/k\gg1
\ee
is fulfilled, a careful analysis shows that:
\be\label{carac1}
w_1(\alpha,\zeta_-) = -\alpha \frac{\zeta_+-\zeta_-}{\mu g}
+ \frac{\alpha^2}{2!} \frac{\zeta_+-\zeta_-}{\mu^3g^2} 
+ O\left( \alpha^3\frac{\zeta_+-\zeta_-}{\mu^5g^3} \right)
\ee
and
\be\label{carac2}
w_2(\alpha,\zeta_-) = 
-\ln\left( 1+\frac{\alpha}{2\mu^2g}\EXP{2\mu(\zeta_+-\zeta_-)} \right)
+\sum_{m=1}^\infty \kappa_m \alpha^m \frac{(\zeta_+-\zeta_-)^m}{g^m\mu^{2m}}
\:.\ee

The characteristic function (\ref{carac1}) corresponds to a sharp Gaussian
distribution $p_1(\Lambda_1)$ for 
\bea
\mean{\Lambda_1}  &=& \frac{\ln(g/k)}{\mu g}\:, \\
\cum{\Lambda_1^2} &=& \frac{\ln(g/k)}{\mu^3g^2}
\:.\eea
This result is not surprising since the potential $U_1$ is monotonic 
and linear on the interesting interval (see figure \ref{potsusy12}), we 
expect the ``time'' to go through the interval to be 
$\frac{\rm distance}{\rm speed}=\frac{\zeta_+-\zeta_-}{\mu g}$ and to 
fluctuate weakly for large enough drift. The precise criterium for the 
validity of this result
is that the time characterising the motion due to the drift,
$\tau_{\rm drift}=\frac{\rm distance}{\rm speed}$, is much shorter than the 
time characterising the motion due to the diffusion,
$\tau_{\rm diff}=\frac{({\rm distance})^2}{\rm diffusion}$; that is
$\frac{(\zeta_+-\zeta_-)}{\mu g}\ll\frac{(\zeta_+-\zeta_-)^2}{g}$, which
leads to (\ref{condition}).

In (\ref{carac2}), the first term gives contributions to the cumulants 
exponentially large in $(\zeta_+-\zeta_-)$ whereas the sum gives 
contributions powers of $(\zeta_+-\zeta_-)$. Then we can forget the second
term and the characteristic function is the logarithm of the Laplace 
transform of a Poisson law 
$p_2(\Lambda_2)=\frac{1}{\smean{\Lambda_2}}\exp-\Lambda_2/\smean{\Lambda_2}$ 
with 
\be
\mean{\Lambda_2} \simeq \frac{1}{2\mu^2g}\EXP{2\mu(\zeta_+-\zeta_-)}
=\frac{1}{2\mu^2g}\left(\frac{g}{k}\right)^{2\mu}
\:.\ee
The fact that the distribution $p_2(\Lambda_2)$ is a Poisson law was expected
from the shape of potential $U_2(\zeta)$ (see figure \ref{potsusy12}):
the potential possesses a well that is able to trap the variable $\zeta$ for 
a long time. 
We can now compare the result of the approximation we have made for 
$\mean{\ell}=\mean{\Lambda_1+\Lambda_2}$ with the exact result 
(\ref{IDoSsusy}). We have found $\mean{\ell}\simeq\frac{\ln(g/k)}{\mu g}
+\frac{1}{2\mu^2g}\left(\frac{g}{k}\right)^{2\mu}
\simeq \frac{1}{2\mu^2g}\left(\frac{g^2}{E}\right)^\mu$
whereas (\ref{IDoSsusy}) gives
$N(E)^{-1}\simeq\frac{1}{2g}\frac{\pi^2}{\sin^2\pi\mu}
\left(\frac{4g^2}{E}\right)^\mu$. Despite the pre-factors being different,
our approximation is indeed able to give the well-known power law behaviour 
of the integrated density of states \cite{OvcEri77,BouComGeoLed90}: 
$N(E)\sim E^\mu$. 
We may also have used the Arrhenius formula (\ref{arrhenlaw}) to find 
$\mean{\Lambda_2}\simeq\frac{\pi}{\mu g}\left(\frac{2\mu g}{k}\right)^{2\mu}$, 
which presents still a different pre-factor (equation (\ref{arrhenlaw}) does 
not give the correct pre-factor maybe because the potential $U_2(\zeta)$ is 
not smooth enough in the neighbourhood of its local minimum in the limit 
$k\to0$ as it is assumed to derive the Arrhenius law (\ref{arrhenlaw})). 
However we may distinguish two levels of approximation in 
what we have done. ({\it a}) We have shown that the distribution of 
$\Lambda_2$ is Poisson, that could be demonstrated in a more general way 
following the proof presented in appendix \ref{escapetime} for the case 
of a potential possessing a local minimum. ({\it b})~We have given an 
approximative expression of the average time $\mean{\Lambda_2}$ that only 
gives the correct behaviour with $E$ but not the correct pre-factor.

However we can avoid the not absolutely satisfactory approximation ({\it b})
because we know that $\mean{\Lambda_1+\Lambda_2}=N(E)^{-1}$ is exact. Then 
we may give the distribution of the length $\ell$ between nodes of the wave 
function: 
\be
P(\ell) = \int\D\Lambda_2\,p_2(\Lambda_2)\,p_1(\ell-\Lambda_2)
\:;\ee
we have shown that $p_2(\Lambda_2)$ varies on a characteristic scale
$\mean{\Lambda_2}\gg\mean{\Lambda_1}\gg\sqrt{\cum{\Lambda_1^2}}$ compared
to which $p_1(\Lambda_1)$ is very narrow. It 
follows that $P(\ell)\simeq p_2(\ell)$ apart from the behaviour at the 
origin we are not interested in, since it is associated with small samples.
Then we conclude that
\be
P(\ell) = N(E) \, \EXP{-\ell\,N(E)}
\:.\ee

We can now give the distribution of the $n$-th excited state which has the 
same form than what was found for the model {\bf A} (\ref{TheResult}). Then we 
have:
\be\label{Wsusymu}
W_n(E) = L\mu a_\mu E^{\mu-1}\frac{(L a_\mu E^\mu)^n}{n!}
\EXP{-L a_\mu E^\mu}
\:,\ee
where the coefficient $a_\mu$ is defined by: $N(E)\simeq a_\mu E^\mu$.
We recall that this result is valid if (\ref{condition}) is fulfilled, that
is $\mu\ln(g/k)\gg1$, in addition to the fact that $k\ll g$.

It follows that the distribution involves the scaling function:
\be
W_n(E) = \frac{1}{\Dv}\ \omega_n\left(\frac{E}{\Dv}\right)
\ee
where
\be
\omega_n(X) = \mu X^{\mu-1} \frac{X^{n\mu}}{n!} \EXP{-X^\mu}
\:,\ee
the energy scale being 
\be
\Dv = \frac{1}{(a_\mu\,L)^{1/\mu}}
\:.\ee
We may easily compute the moments:
\be
\mean{E_n^m} = \Dv^m \frac{\Gamma(n+1+m/\mu)}{\Gamma(n+1)}
\:.\ee

It is also interesting to compare the mean level spacing 
$\mean{\Delta_n}=\mean{E_{n+1}-E_n}$ 
and the fluctuations $\delta E_n=\sqrt{\cum{E_n^2}}$. We get:
\be
\mean{\Delta_n}=\Dv\frac{\Gamma(1/\mu+n+1)}{\mu\,(n+1)!}
\:.\ee
Then 
\be
\frac{\delta E_n}{\mean{\Delta_n}} = \mu\,(n+1)
\sqrt{ n! \frac{\Gamma(2/\mu+n+1)}{\Gamma(1/\mu+n+1)^2} - 1 }
\:,\ee
which becomes large at small $\mu$ ($\ll1/n^2$):
\be
\frac{\delta E_n}{\mean{\Delta_n}} \propto \mu^{n/2+5/4} 2^{1/\mu}
\:.\ee
At large $\mu$ the ratio reaches a constant value:
\be
\lim_{\mu\to\infty}\frac{\delta E_n}{\mean{\Delta_n}} 
=(n+1)\sqrt{\psi'(n+1)}
\:,\ee
where $\psi(z)$ is the digamma function (in particular $\psi'(1)=\pi^2/6$). 
For small $n$ the fluctuations and the mean level spacing are of same order.

\vspace{0.25cm}

The distribution (\ref{Wsusymu}) has the form 
of the distribution of the $n+1$-th lowest variable among $N\to\infty$ 
statistically independent variables distributed by a law behaving like 
$p(E)\sim E^{\mu-1}$ (see appendix \ref{extreme}), the same behaviour as
the density of states. This remark shows that the energies behave like 
statistically independent ordered variables, in agreement with the expected 
absence of level repulsion in the localized regime.


\section{Conclusion}

We have considered two one-dimensional disordered models: one with diagonal
disorder ({\bf A}) and another with off-diagonal disorder ({\bf B}). 

For model {\bf A} we have derived the distribution of the distance between 
consecutive nodes of the wave function in the limit $|E|\gg\sigma^{2/3}$
both in the negative part of the spectrum and in the positive part.
Using these results we were able to find the distribution of the $n$-th 
excited state (\ref{TheResult},\ref{Wn+}). For $E<0$ we have shown that
(\ref{TheResult}) is a scaling law (\ref{TheResultbis}), similar to the 
extreme statistics of independent variables.
If $E<0$ the typical value of the $n$-th energy
behaves with the size of the system like: 
$E_n^{\typ}\sim-\ln^{2/3}L$. The width of its distribution behaves like
$\delta E_n\sim\ln^{-1/3}L$. If $E>0$ we have found
$E_n^{\typ}\sim1/L^2$ and $\delta E_n\sim1/\sqrt{L}$. Note however that 
the relative fluctuations of $E_n$ in this latter case are small despite
the behaviours with $L$ suggest the opposite.

For model {\bf B} we have first considered the case $\mu=0$ (mean value
of the function entering the supersymmetric potential). The high energy
limit (universal regime) gives results similar to those for model {\bf A}.
In the low energy limit we have found the distribution for the ground 
state (\ref{W0susy},\ref{pi0susy}). We have shown that this distribution is 
broad, its positive moments being all dominated by the tail of the 
distribution:
$E_0^{\typ}\sim\EXP{-L}\ll E_0^{\rm med}\sim\EXP{-\sqrt{L}}
\ll \mean{E_0}\sim\EXP{-L^{1/3}}$. 
The moments have an atypical $n$-dependence: $\smean{E_0^n}\sim b^{n^{2/3}}$.
We have also given explicitely the distribution for the second energy level
(\ref{om1susy},\ref{om1susyB}) and an integral representation for the other 
energy levels (\ref{pinsusy}).
For $\mu=0$ these distributions do not have
the form of the distribution of extremes of independent variables.
For $\mu\neq0$ we were able to derive the distributions $W_n(E)$ in the 
low energy limit (in the high energy limit, universal regime, we do not 
expect any difference with the picture obtained for $\mu=0$). We have shown 
that $W_n(E)$ is a scaling function $\omega_n(E/\Dv)$ where the energy scale 
behaves like:
$\Dv\propto L^{-1/\mu}$. For $\mu\neq0$ the distribution exhibits 
extreme value statistics as for model {\bf A}.

We now discuss the relation between the distributions $W_n(E)$ 
we have found and the extreme value statistics (see appendix \ref{extreme}). 
We have seen that for model {\bf A} and for model {\bf B} with $\mu\neq0$, 
$W_n(E)$ has the form of a distribution of extreme values.
If we suppose that energies are behaving like statistically independent 
random variables, the distribution of one of them should be proportional to 
the density of states; 
the number ${\cal N}$ of these variables is proportional to the 
size of the system. If we replace ${\cal N}$ by $L\times\Omega$ and $p(x)$ 
by $\rho(E)/\Omega$ in (\ref{extreq}), where $\Omega$ is the total
number of states per unit length (in principle infinite since the spectrum 
is unbounded from above), it is easy to see that we get the equation 
(\ref{TheResult}) obtained both for model {\bf A} and for model {\bf B} 
with $\mu\neq0$. 
We now consider more specifically the case of the ground state ($n=0$), since
$W_0(E)$ is directly related to the distribution $P(\ell)$, and proceed in the
opposite way we have followed until now. If we admit that the distribution of 
the lowest energy is $W_0(E)=L\rho(E)\exp-L\,N(E)$ (see equation 
(\ref{TheResult})),
as a consequence of the statistical independence of the energies, then we 
conclude that $P(\ell)$ is Poissonian. Conversely if $P(\ell)$ is not Poisson
in the range of energy where $E_0$ is expected to be found, there might exist
some correlations between the energies.
This is indeed the case for the high energy limit: the 
distribution (\ref{Wn+}) for $n=0$, a consequence of the narrow Gaussian 
distribution $P(\ell)$, is valid only in the delocalized regime due to 
condition (\ref{validity}), and in this regime level repulsion occurs, as 
explained, due to (\ref{LRornot}). 
This argument can only be used for the ground state energy 
distribution since $W_n(E)$ is not directly related to $P(\ell)$ in 
this case. So the fact that (\ref{Wn+}) for $n\gg1$ comes from a Gaussian
narrow distribution $P(\ell)$ in this range of energy does not necessarily 
mean level repulsion.
Since the distribution of the ground state for the supersymmetric model with 
$\mu=0$ is $W_0(E)=L\rho(E)\varpi_0(L\,N(E))$ with $\varpi_0(X)\neq\EXP{-X}$, 
this suggests that level 
repulsion might occur at the bottom of the spectrum. This seems
to contradict the expected absence of level repulsion in localized regime
since $W_0(E)$ shows that the most probable energy 
$E\gg E^{\typ}\simeq g^2\exp-gL$, 
{\it i.e.} $\lambda\simeq\frac1g\ln(g^2/E)\ll L$, are associated with 
localized states. 
However $\lambda\simeq\frac1g\ln(g^2/E)$ is the localization length for the 
infinite size system, given by the inverse of the mean Lyapunov exponent 
$\mean{\gamma}=1/\lambda$ which is going to zero if $E\to0$. For finite size 
system the Lyapunov exponent has some Gaussian fluctuations which cause 
fluctuations of the localization length: $\cum{\gamma^2}\propto1/L$.
For a given $L$ the fluctuations of the Lyapunov exponent become of the order
of its mean value when the energy becomes smaller than an energy 
$g^2\exp(-c\sqrt{gL})$ of the order of the median value of the distribution
$W_0(E)$; in this case, the large fluctuations of the Lyapunov exponent may 
cause delocalization and it might not be surprising that level correlation
appears in that regime being the reason of the fact that $W_0(E)$ does not
behave like the extreme value distribution of independent variables as it 
is the case for $\mu\neq0$ or for model {\bf A}.
Nevertheless a deeper understanding of level correlations seems to be needed 
for the supersymmetric case with $\mu=0$.

Since model {\bf B} is related to the problem of classical diffusion
in a random medium, it would be interesting to know if these results 
have an application in this case. The energies of the Hamiltonian 
should be related to relaxation times \cite{LedMonFis99}. 
The spectral properties near $E=0$ are important for the disordered
spin chain models \cite{SteFabGog98}, the distribution $W_0(E)$ of the 
lowest mode of the Hamiltonian (\ref{Hsusy}) might also have some interest 
in that context.


\section*{Acknowledgments}

I am grateful to Alain Comtet for having motivated my interest in this 
problem and with whom I had very interesting discussions (in particular
he brought to my attention Refs. \cite{Gum54,Gum58,Gal78}).
I thank Gleb Oshanin for interesting remarks. I also had stimulating 
discussions with Marc Bocquet and St\'ephane Ouvry.
This work was supported by the Swiss National Science Foundation 
and by the TMR Network Dynamics of Nanostructures. 


\begin{appendix}

\section{Distribution of the escape time for a diffusive particle trapped
         in a well \label{escapetime}}

In this appendix we recall known results about the trapping of a Brownian 
particle by a well and the distribution of the escape time. In the 
regime of interest here the average escape time is given by the Arrhenius
law. Most of what is in this appendix may be found in standard textbooks 
like \cite{Gar89} and we summarize here some ideas needed throughout this 
article. 

We consider a particle whose position $x(t)$ obeys the following 
stochastic differential equation:
\be
\D x(t) = A(x) \D t + \sqrt{B(x)}\:\D W(t) \hspace{1cm} ({\rm Ito})
\ee
where $W(t)$ is a normalized Wiener process: 
$\mean{W(t)}=0$ and $\mean{W(t)W(t')}=\min{t}{t'}$.
This equation is understood
in the Ito sense. The propagator $p(x',t|x,0)$ for the diffusion
(conditional probability for the particle to be at $x'$ at time $t$, 
starting from $x$ at initial time $0$) obeys the backward Fokker-Planck 
equation (BFPE):
\be
\partial_t p(x',t|x,0) = G_{x} \, p(x',t|x,0)
\:,\ee
where 
\be
G_{x} = A(x) \partial_{x} + \frac12 B(x) \partial_{x}^2
\ee
is the BFPE generator.

Let us consider a process starting from a point $x(0)=x$ belonging to the 
interval $[a,b]$. Let us call $T$ the time needed by the particle
to leave this interval. The probability that the particle is still in the 
interval after time $T$ is 
\be\label{rel1}
\int_a^b\D x'\, p(x',T|x,0) = \int_T^\infty\D T'\,P_x(T')
\:,\ee
where $P_x(T)$ is the probability density for the escape time.
Let us define $T_n(x)=\smean{T^n \:|\: x(0)=x;\: x(T)=a\ {\rm or}\ b}$, the 
$n$-th moment of the escape time from the interval $[a,b]$. It follows from 
(\ref{rel1}) that 
\be\label{eqdiffTn}
G_x \, T_n(x) = -n T_{n-1}(x)
\:.\ee
Since the particle may escape from both sides of the interval, we have to 
impose the following boundary conditions: $T_n(a)=T_n(b)=0$. This absorbing
boundary conditions mean that the particle leaves the interval and never 
comes back as soon as it reaches one of the two edges. In this sense we are 
interested in a time of first exit, relevant for the question 
considered in this article.

In the following we will be interested in the more simple case of
a reflection condition at one side of the interval, $x=a$, so that the 
particle may only escape from the side $x=b$: 
\be
T_n(x)=\smean{T^n \:|\: x(0)=x;\: x(T)=b}
\:.\ee
The reflecting boundary 
condition for the BFPE is $\partial_xp(x',t|x,0)|_{x=a}=0$ \cite{Gar89}
which implies the following boundary conditions for the moments:
\bea
\partial_x T_n(a) &=& 0 \\
T_n(b) &=& 0
\:.\eea
In this case it is easy to construct from (\ref{eqdiffTn}) the moments:
\be\label{recumom}
T_n(x) = 2n \int_x^b   \D x' \,\frac{1}{\psi_0(x')}
            \int_a^{x'}\D x''\,\frac{\psi_0(x'')}{B(x'')}\:T_{n-1}(x'') 
\ee
where 
\be
\psi_0(x)=\exp\int^x\D x'\, \frac{2 A(x')}{B(x')}
\:.\ee
Keeping in mind that $T_0(x)=1$, it follows that the moments may be 
computed recursively.

We also introduce the generating function for the distribution $P_x(T)$:
\be
h(\alpha,x)=\smean{\EXP{-\alpha T} \:|\: x(0)=x,\: x(T)=b}
\:,\ee
which may be used to analyze the distribution. It is clear from 
(\ref{eqdiffTn}) that it obeys the following differential equation:
\be\label{eqgenfu}
G_x \, h(\alpha,x) = \alpha \, h(\alpha,x)
\:.\ee
The boundary conditions for reflection in $a$ and escape (absorption) at $b$ 
are:
\bea\label{b1}
\partial_x h(\alpha,a) &=& 0 \\
\label{b2}h(\alpha,b) &=& 1
\:.\eea
Instead of considering (\ref{eqgenfu},\ref{b1},\ref{b2}) it can be more 
convenient to write an integral equation for $h$:
\be\label{eqinth}
h(\alpha,x) = 1 - 2 \alpha\int_x^b   \D x' \,\frac{1}{\psi_0(x')}
            \int_a^{x'}\D x''\,\frac{\psi_0(x'')}{B(x'')}\:h(\alpha,x'')
\:.\ee

\begin{figure}[!ht]
\begin{center}
\includegraphics[scale=0.9]{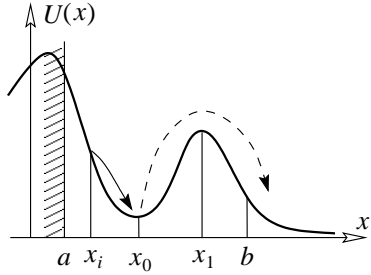}
\end{center}
\caption{The particle starts from $x_i$ and escape the well after a time
$T$.\label{fig:Arrhenius}}
\end{figure}

We now derive the distribution $P_x(T)$ when the particle is trapped by the  
well of a potential $U(x)$, in the small noise limit. We consider the more 
simple situation of a stochastic differential equation of the form:
\be
\D x(t) = -U'(x) \D t + \sqrt{D}\:\D W(t) 
\ee
where $D$ is the diffusion constant.  The shape of the potential of interest 
is depicted in figure
\ref{fig:Arrhenius}. The potential $U(x)$ has a local minimum $x_0$ from
the neighbourhood of which the particle starts. Let us call $x_i$ the initial
position of the particle. The potential has at $x_1>x_0$ a local maximum. 
We are interested in the time that the particle need to escape the well, 
jumping over the potential barrier due to a fluctuation. The interval 
$[a,b]$ to be considered has to include the well and the barrier. As
it will be clear in the following we have to consider a limit $b$ such that 
the distance $b-x_1$ is sufficiently large (the relevant length scale is 
given by the curvature of $U(x)$ at $x_1$). For similar reasons the 
distance $x_0-a$ has to be sufficiently large as well. Apart from this 
two restrictions the precise positions of $a$ and $b$ are of little 
importance.

Let us analyze the behaviour of the moments when the diffusion $D$ is small 
compared to the height of the barrier. 
Equation (\ref{recumom}) now takes the form:
\be\label{recumom2}
T_n(x_i) = \frac{2n}{D}
\int_{x_i}^b\D x\,\EXP{2U(x)/D}\int_a^x\D x'\, \EXP{-2U(x')/D} \, T_{n-1}(x')
\:.\ee
From the shape of the potential of figure \ref{fig:Arrhenius}, it is clear
that the integral over $x$ is dominated by the neighbourhood of $x_1$ and 
the integral over $x'$ by the neighbourhood of $x_0$. Accordingly, we have:
\be
T_n(x_i) \simeq \frac{2n}{D} T_{n-1}(x_0)
\int_{x_i}^b\D x\,\EXP{2U(x)/D}\int_a^{x_1}\D x'\, \EXP{-2U(x')/D} 
\:.\ee
Since we are dealing with the limit $D\to0$, the two integrals may be 
estimated by the steepest descent method. Expanding the potential in the 
neighbourhood of its two extrema: 
\bea
&& U(x) \APPROX{x\sim x_0} U(x_0) + \frac{(x-x_0)^2}{2\delta_0^2} \\
&& U(x) \APPROX{x\sim x_1} U(x_1) - \frac{(x-x_1)^2}{2\delta_1^2}
\:,\eea
we eventually find:
\be
T_n(x_i) \simeq n \, T_{n-1}(x_0) \,
2\pi\,\delta_0\,\delta_1\,\EXP{2\frac{U(x_1)-U(x_0)}{D}}
\:.\ee
For $n=1$, we recover the well-known Arrhenius law \cite{Gar89}:
\be\label{arrhenlaw}
T_1(x_0) \simeq 2\pi\,\delta_0\,\delta_1\,\EXP{2\frac{U(x_1)-U(x_0)}{D}}
\:.\ee
Let us remark that for $b=x_1$ we find half of this result.
It is now straightforward to see that the moments are:
\be
T_n(x_0) \simeq n! \, \left[T_1(x_0)\right]^n
\:,\ee
{\it i.e.} the moments of a Poisson law. It follows that the escape time 
from a deep well is distributed by:
\be\label{dtt}
P_{x_i}(T) = \frac{1}{T_1(x_0)} \EXP{-T/T_1(x_0)}
\:.\ee
Let us stress that the initial condition $x_i$ played no role in the 
previous discussion: $T_n(x_i)\simeq T_n(x_0)$.

The same analysis may also be performed using (\ref{eqinth}).


\section{Extreme value statistics \label{extreme}}

We give here a brief discussion on extreme order statistics, details of which 
can be found in \cite{Gum54,Gum58,Gal78}. 

We consider a set of ${\cal N}$ statistically independent variables $x_n$, 
distributed according to the law $p(x)$. Then the ${\cal N}$ variables are 
ordered and we call $w_n(x)$ the probability density of the $n$-th variable 
among these variables. We have:
\be\label{extreq}
w_n(x)\D x = n C_{\cal N}^n \left(\int_{-\infty}^x\D x'\,p(x')\right)^{n-1}
p(x) \D x \left(\int_x^{+\infty}\D x'\,p(x')\right)^{{\cal N}-n}
\:.\ee

We are now interested in the behaviour of this expression in the limit 
${\cal N}\to\infty$. Three kinds of distribution are usually distinguished.

\hspace{0.25cm}

\noindent {\it Type I}.
$p(x)$ is unbounded from below and decays exponentially. 
In the limit ${\cal N}\to\infty$ the distribution $w_n(x)$ is expected
to be peaked around a value $x^{\typ}_n\to-\infty$ so that we may use the 
fact that $p(x)$ may be locally approximated by an exponential in the 
neighbourhood of $x^{\typ}_n$. Then we get, up to a rescaling
$w_n(x)\D x = v_n(y)\D y$:
\be
v_n(y) = \frac{n^n}{(n-1)!} \exp{\left(n y -n\EXP{y}\right)}
\:,\ee
where $y$ is the rescaled variable. The relation between $x$ and the scaling
variable $y$ depends on the distribution $p(x)$ but not the scaling function
$v_n(y)$.

\hspace{0.25cm}

\noindent {\it Type II}.
$p(x)$ has a power law tail. We do not discuss this case which is
not relevant for what is done in this article.

\hspace{0.25cm}

\noindent {\it Type III}.
$p(x)$ is bounded from below. We assume its support starts at $x=0$ 
and it behaves like $p(x)\sim x^{\mu-1}$ for small $x$. Then, up to a 
rescaling of the variable, we end with:
\be
v_n(y) = \frac{n^n\mu}{(n-1)!} y^{\mu\,n-1}\EXP{ -n y^\mu}
\:.\ee

\end{appendix}




\end{document}